\documentclass[iop,apj]{emulateapj}
\usepackage[backref,breaklinks,colorlinks,citecolor=blue]{hyperref}

\usepackage[all]{hypcap} 
\usepackage{ulem}

\begin{document}
\title{X-ray outbursts of ESO 243-49 HLX-1: comparison with Galactic low-mass X-ray binary transients}
\author{ Zhen Yan\altaffilmark{1},   Wenda Zhang\altaffilmark{1}, Roberto Soria
\altaffilmark{2}, Diego Altamirano\altaffilmark{3} and Wenfei Yu\altaffilmark{1} }
\altaffiltext{1}{Key Laboratory for Research in Galaxies and Cosmology, Shanghai 
Astronomical Observatory, Chinese Academy of Sciences, 80 Nandan Road, Shanghai 200030, 
China;  \url{zyan@shao.ac.cn } }
\altaffiltext{2}{International Centre for Radio Astronomy Research, Curtin University, GPO 
Box U1987, Perth, WA 6845, Australia}
\altaffiltext{3}{Physics and Astronomy, University of Southampton, Southampton, Hampshire 
SO17 1BJ, UK}

\begin{abstract}
We studied the outburst properties of the hyper-luminous X-ray source ESO\,243-49 HLX-1, 
using the full set of {\it Swift} monitoring observations. We quantified the increase in the 
waiting time, recurrence time, and $e$-folding rise timescale along the outburst sequence, and the corresponding 
decrease in outburst duration, total radiated energy, and $e$-folding decay timescale, 
which confirms previous findings. HLX-1 
spends less and less time in outburst and more and more time in quiescence, but its peak 
luminosity remains approximately constant. We compared the 
HLX-1 outburst properties with those of bright Galactic low-mass X-ray binary transients 
(LMXBTs). Our spectral analysis strengthens the similarity between state transitions in HLX-1 and 
those in Galactic LMXBTs. We also found that HLX-1 follows the nearly linear correlations between the
hard-to-soft state transition luminosity and the peak luminosity, and between the rate of change of X-ray luminosity 
during the rise phase and the peak luminosity, which indicates that the occurrence of 
the hard-to-soft state transition of HLX-1 is similar to those of Galactic LMXBTs during outbursts.
We found that HLX-1 does not follow the correlations between total radiated energy 
and peak luminosity, and between total radiated energy and $e$-folding rise/decay timescales we 
had previously identified in Galactic LMXBTs. 
HLX-1 would follow those correlations if the distance were several hundreds of kiloparsecs. 
However, invoking a much closer distance for HLX-1 is not a 
viable solution to this problem, as it introduces other, more serious inconsistencies with the 
observations. 
\end{abstract}

\keywords{accretion, accretion disks --- black hole physics ---X-rays:binaries --- X-rays: 
individual (HLX-1)}

\section{INTRODUCTION}
HLX-1 is the most luminous ultraluminous X-ray source \citep[ULX;][for a review]
{Feng2011} known with a maximum isotropic X-ray luminosity of $> 10^{42}~ 
\mathrm{erg~s^{-1}}$ \citep{Farrell2009,Godet2009}. The extremely high luminosity makes 
HLX-1 currently the best candidate to harbor an intermediate-mass black hole (IMBH, 
10$^{2}$--10$^{5}$ M$_{\sun}$) among ULXs. Although other ULXs in nearby galaxies were initially 
flagged as candidate IMBHs, recent works show that some of them contain a stellar-mass black 
hole \citep[BH;][]{Liu2013, Motch2014} or even a neutron star \citep[NS;][]{Bachetti2014}. HLX-1 is 
projected in the sky within the D25 (the diameter of the light isophote having a surface 
brightness of 25.0 mag arcsec$^{-2}$) of the S0a spiral galaxy ESO\,243-49 (redshift distance 
$\approx$95 Mpc), $\approx8\arcsec$ north-east of its nucleus \citep{Farrell2009}. The 
redshift measurement of the H${\alpha}$ emission line from the optical counterpart of HLX-1 
confirmed the association of HLX-1 with ESO\,243-49 \citep{Wiersema2010,Soria2013}. 
However, there is an offset of $\approx$430 km s$^{-1}$ between the recession velocity of 
HLX-1 and that of the nucleus of ESO\,243-49: this suggests that HLX-1 may be in a dwarf 
satellite galaxy or star cluster near ESO\,243-49 rather than in the galaxy itself 
\citep{Soria2013}.

Monitoring observations of HLX-1 with {\it Swift}/XRT have shown that the source undergoes 
recurrent outbursts (almost once a year), where the outburst profiles are well described by a 
fast-rise-exponential-decay (FRED) form \citep[see also][]{Lasota2011,Godet2014}, similar to 
the outbursts of many bright Galactic low-mass X-ray binary transients
\citep[LMXBTs; see ][]{Chen1997,Yan2015}. During its outbursts, HLX-1 displayed spectral 
state transitions which also resemble those of the bright Galactic LMXBTs 
\citep{Godet2009,Servillat2011}. \citet{Webb2012} have detected transient radio emission 
following the hard-to-soft state transition during the 2010 and 2011 outbursts. By analogy with 
the radio flares seen in Galactic BH LMXBs  \citep{Fender2004}, the 
transient radio emission was interpreted as discrete jet ejection events during the state transitions 
\citep{Webb2012}.

It is generally believed that the outbursts of bright Galactic LMXBTs are triggered by 
the thermal-viscous disk instability \citep[][for a review]{Lasota2001}. In this model, the outburst duration 
is approximately the viscous timescale at the minimum unstable disk radius, which is the radius at 
which the irradiated disk becomes cold enough to be dominated by neutral hydrogen. If HLX-1 
is located at $\approx$95 Mpc, then its maximum luminosity, $\approx10^{42}$ erg s$^{-1}$,  
implies a viscous timescale $\sim$100 years \citep[see][assuming that the disk is large 
enough to allow for a neutral outer region]{Lasota2011,Lasota2015}. However, the observed outburst 
duration of HLX-1 is $\sim$100 days, i.e., at least 2 orders of magnitude shorter than needed, 
therefore suggesting that the outbursts are triggered from a much smaller radius than required by 
the thermal-viscous disk instability model. 
Clearly, another mechanism is probably needed to 
trigger the outbursts.  For example, \citet{Lasota2011} proposed that outbursts are triggered by 
the passage at periapse of the companion star on an eccentric orbit. However, this model cannot 
easily explain why the outburst repeats at irregular intervals (between $\approx$370 and 460 
days: \citealt{Kong2011, Lasota2011, Godet2013, Godet2014, Godet2015,  Miller2014, Kong2015}). Based on the observed decline of the peak count rate, integrated luminosity, duration, and decay time over the sequence of outbursts, it was also suggested that the donor star may be undergoing tidal disruption 
\citep{Godet2013, Godet2014, Miller2014}. \citet{Miller2014} proposed an alternative model 
based on (low-angular-momentum) wind accretion instead of Roche lobe overflow. 
\citet{King2014} suggested that HLX-1 may be a stellar-mass binary system like SS\,433
\citep[see also][]{Lasota2015}, in 
which the X-ray emission comes from the beamed jet. None of those scenarios appear to be fully consistent with all the observational constraints, and there is still no general consensus on what causes the outbursts.

In previous work \citep{Yan2015}, we have systematically studied the outburst properties of 
bright Galactic LMXBTs, quantifying statistical correlations between observable quantities and 
comparing them with the predictions of the thermal-viscous disk instability model. In order to 
understand the nature of HLX-1 outbursts, in this paper, we use the same parameters as used by 
\citet{Yan2015} to quantify the outburst properties of HLX-1, how they changed from outburst 
to outburst, and how they compare with the observed properties of bright Galactic LMXBTs 
outbursts.

\section{DATA ANALYSIS AND RESULTS}

\subsection{Different Spectral States}
\label{sec2.1}
\begin{figure}  
\centering    
\includegraphics[width=\linewidth]{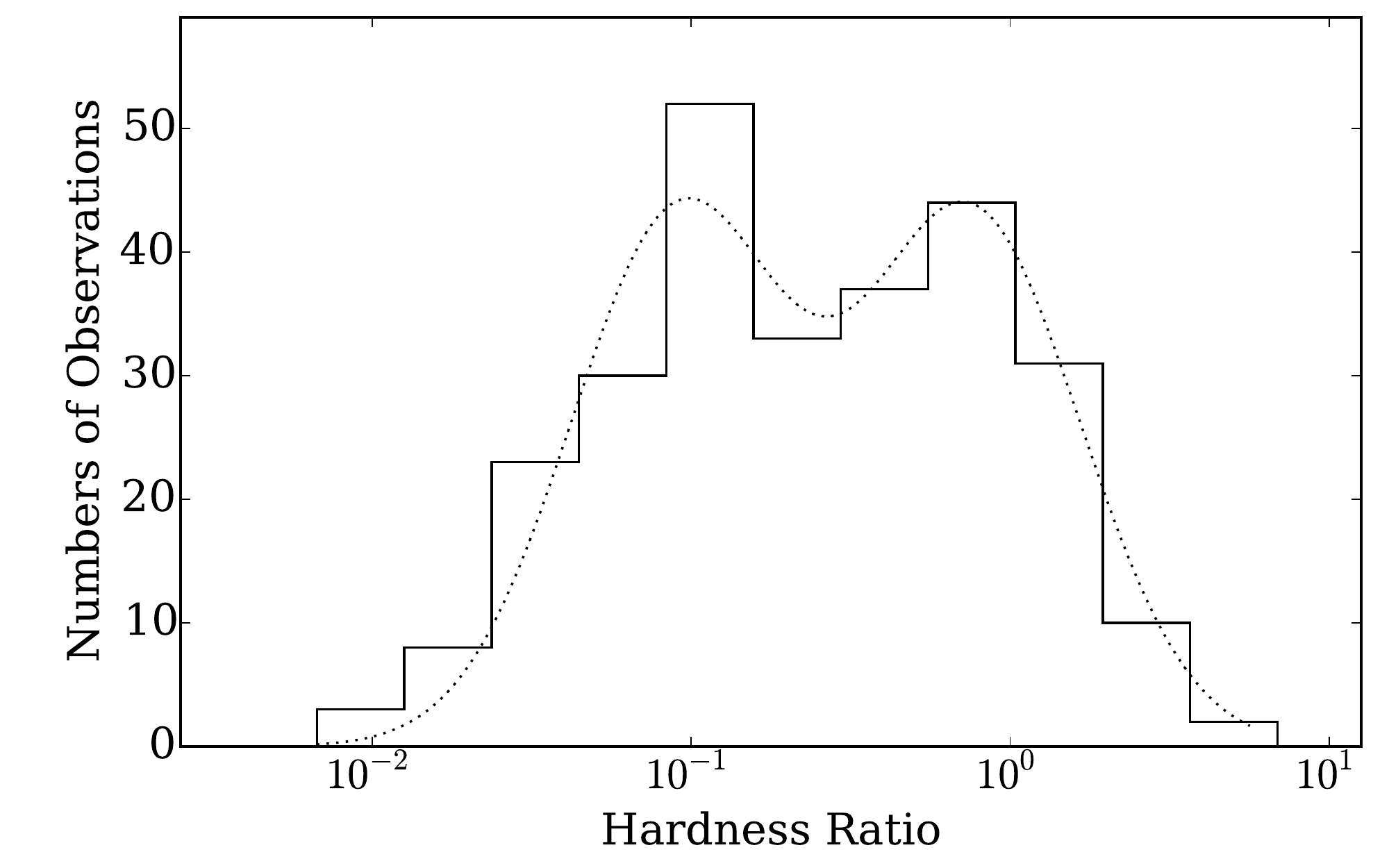}
\caption{Histogram of the (1.5--10 keV)/(0.3--1.5 keV) HR distribution
  (solid line), fit with a double Gaussian model (dotted line). The
  Gaussian peaks are roughly at HR $\approx$ 0.1 and HR $\approx$ 0.6.}
  \vspace{0.5cm}
\label{hr}
\end{figure}

The {\it Swift} X-Ray Telescope (XRT) has monitored HLX-1 regularly for  more than six years, 
providing us with an opportunity to investigate the X-ray temporal and spectral variability of 
this source. We obtained the $0.3$--$10$ keV count rate and the (1.5--10 keV)/(0.3--1.5 keV) 
hardness ratio (HR) of each {\it Swift}/XRT observation from the online {\it Swift}/XRT 
product generator \citep{Evans2007,Evans2009}. We plotted the HR distribution from all 
observations with positive HR (\autoref{hr}): it shows two peaks, suggesting that HLX-1 has 
two distinct spectral  states (hard state and soft state) by analogy with Galactic X-ray binaries \citep[XRBs;][]{Yu2009,Yan2010}. 
The presence of different spectral states in HLX-1 was first noted by 
\cite{Godet2009} and was also investigated by \citet{Servillat2011}. We used the double-peaked HR 
distribution to define the hard and soft states: we fitted the histogram with two Gaussian 
components (\autoref{hr}) and found that the centroids are roughly at HR $\approx 0.1$ and HR 
$\approx 0.6$. We classify all the observations with HR $\le0.1$ as in the soft state, and all 
those with HR $\ge 0.6$ as in the hard state. The observations corresponding to 
$0.1 < {\mathrm{HR}} < 0.6$ consist of a mixture of soft, hard and perhaps 
intermediate states: we classify those observations as the transitional state.

\capstartfalse
\capstarttrue

\begin{figure}  
\centering    
\includegraphics[width=\linewidth]{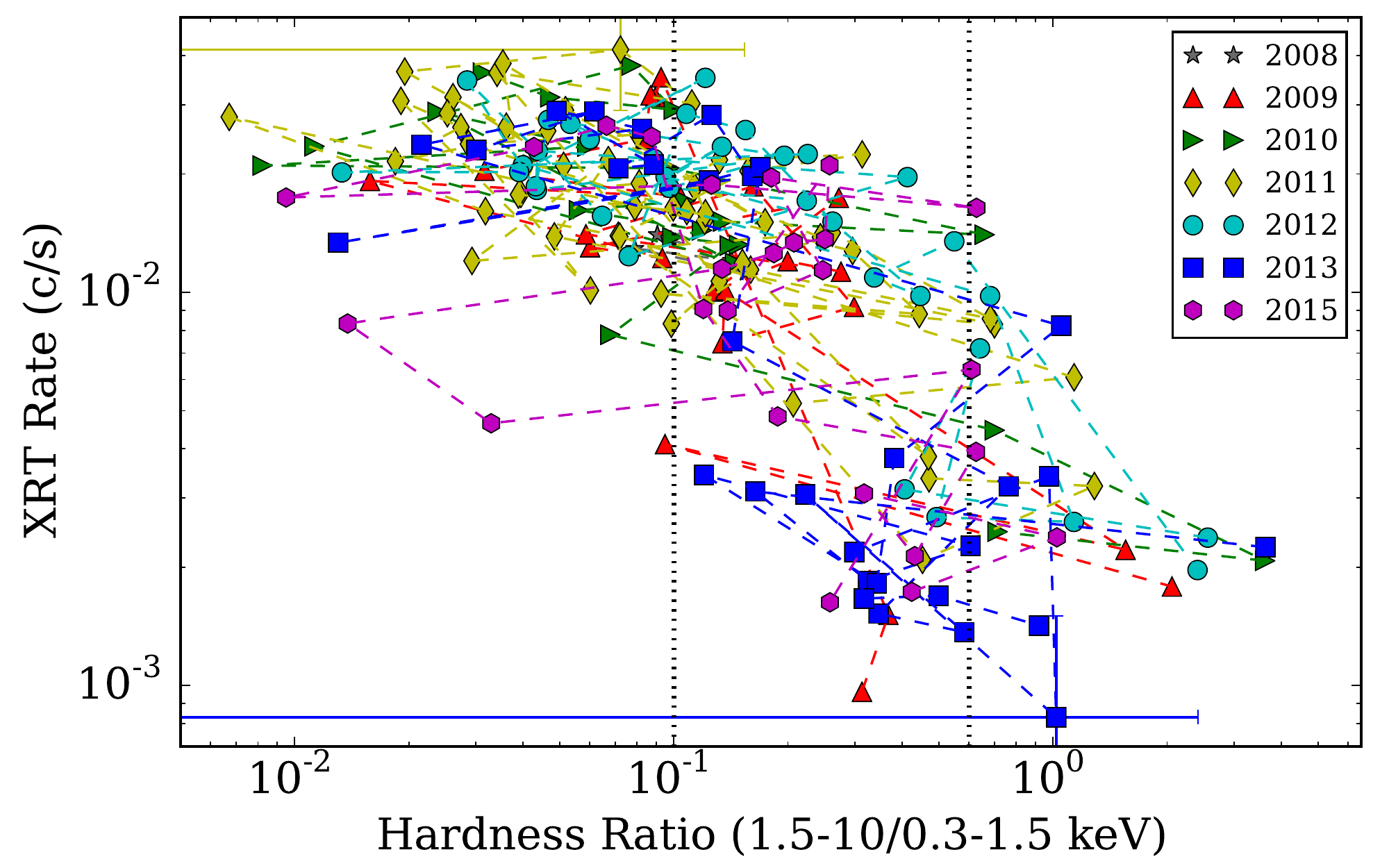}
\caption{$Swift$/XRT hardness$-$intensity diagram of HLX-1. We include all observations with 
detection significance $>2\sigma$ in the full 0.3--10 keV band \citep{Evans2007,Evans2009}. 
For clarity, we only plot two points with error bars to indicate representative uncertainty levels. The 
dotted lines separate the soft, transitional and hard spectral states.  For each outburst, dashed 
lines indicate the spectral evolution from the hard to the soft state and then back to the hard 
state.}
\vspace{0.5cm}
\label{hid}
\end{figure}

Hardness$-$intensity diagrams (HIDs) are a very useful tool to characterize the different spectral 
states in Galactic LMXBs \citep[\textit{e.g.}][]{Miyamoto1995, Homan2001,Maccarone2003a, 
Kording2008, Belloni2010}.  In \autoref{hid} we plot the HID of all HLX-1 outbursts. The 
diagram clearly shows two clusters of data points, corresponding to the high/soft and low/hard state. In each 
outburst, HLX-1 evolved from the hard state to the soft state during the rise phase and then returned 
to the hard state during the decay phase. This behavior is similar to most outbursts of Galactic
LMXBTs \citep[\textit{e.g.}][]{Homan2001,Belloni2006,Belloni2010,Dunn2010,Munoz-Darias2014}.

\capstartfalse
\capstarttrue

\begin{figure}  
\centering    
\includegraphics[width=\linewidth]{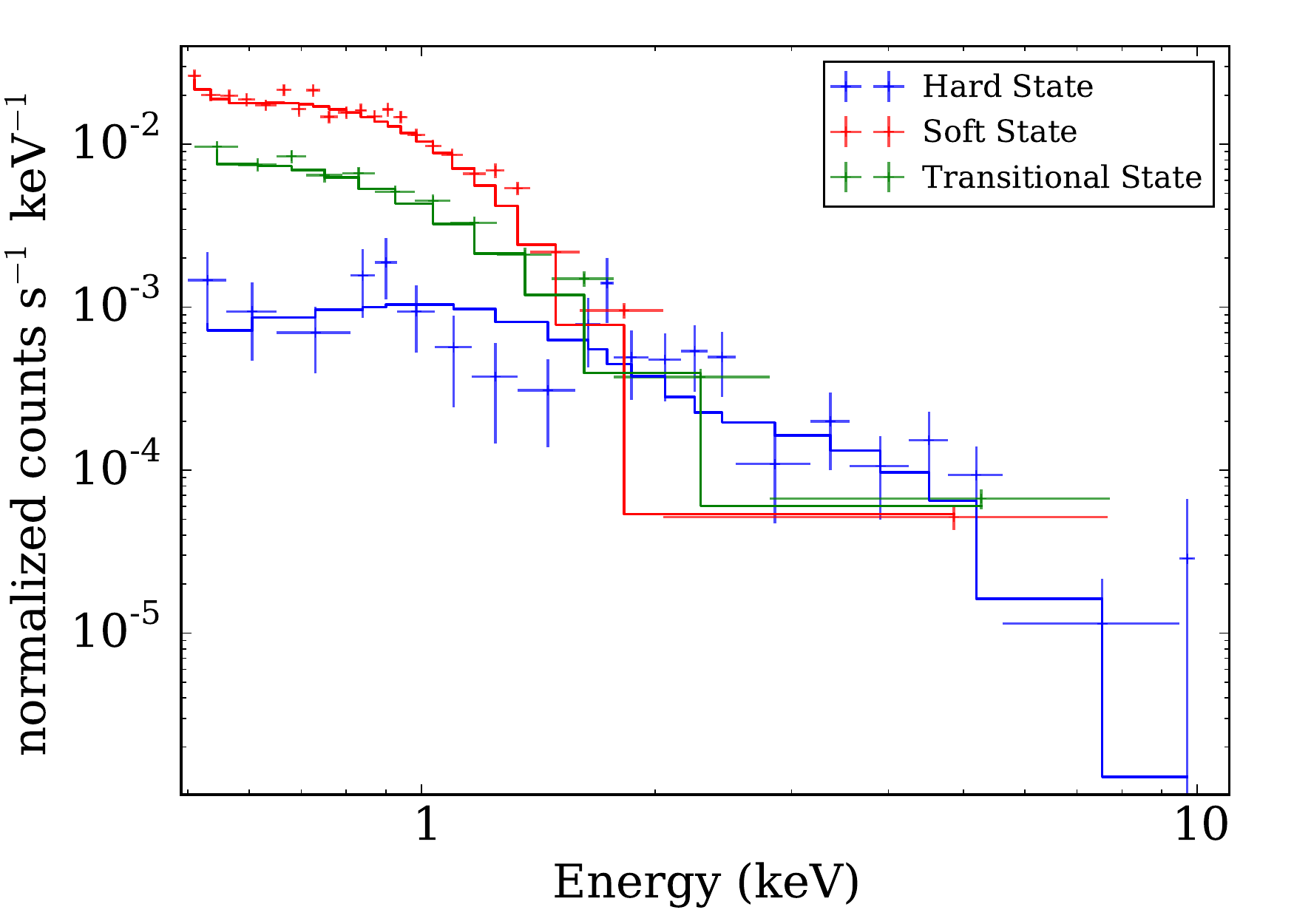}
\caption{Stacked {\it Swift}/XRT spectra and best-fitting models for
  the three states defined in \autoref{sec2.1}: red data points for the soft
  state, green for the transitional state, and blue for the hard
  state. See \autoref{fit} for the best-fitting parameters. }
  \vspace{0.5cm}
\label{spectra}
\end{figure} 

\capstartfalse
 \begin{deluxetable*}{lccccl}
\centering
\tablecaption{Best-fitting parameters in different spectral states}
\tablewidth{0pt}
\tablehead{
\colhead{Spectral States} &\colhead{$T_{\rm in}$ (keV)}      & \colhead{$N_{\rm disk}$}   &
\colhead{$\Gamma$} & \colhead{$N_{\rm pow}~(10^{-5})$}  &\colhead{$\chi^2$/dof} }
\startdata
Hard State & \nodata & \nodata &$1.62^{+0.26}_{-0.32}$ &1.60$^{+0.37}_{-0.36}$ & 67.27/104 (C-statistic)\\
Transitional State &$0.16^{+0.02}_{-0.03}$ &25.60$^{+32.83}_{-12.06}$ &2.45$^{+0.41}_{-0.45}$ &3.62$^{+1.51}_{-1.36}$ &  38.26/52 \\
Soft State &0.20$^{+0.01}_{-0.01}$ & $32.08^{+7.65}_{-7.12}$ &$2.56^{+0.75}_{-0.82}$ &$2.13^{+2.20}_{-1.33}$  & 81.52/82 \\
\enddata
\label{fit}
\end{deluxetable*}
\capstarttrue

We extracted stacked spectra of the {\it Swift}/XRT observations in each of the three states 
(\autoref{spectra}), using the online XRT product generator \citep{Evans2009}, which also 
provides the associated background and response files. We only included observations with 
detection significance $> 2\sigma$. We analyzed and modeled the spectra with 
{\small{XSPEC}} Version 12.8.2 \citep{Arnaud1996}.  Because of the low count rate, 
we grouped the spectrum of the hard state to at least 1 count bin$^{-1}$, and applied the Cash 
statistics \citep{Cash1979}, and grouped the spectra of the transitional and soft sates to at least 20 
counts bin$^{-1}$, so that we could use $\chi^{2}$ statistics. 
We ignored photon energies below 0.5 keV. 
We have tried to free the column density, but it can only be well constrained by the spectrum of
the soft sate at a value of $N_{\rm H} = 5^{+3}_{-2}\times10^{20}$ cm$^{-2}$, which is
consistent with the value used before \citep{Farrell2009,Webb2010,Webb2012}. Therefore, we fixed the
column density at the line-of-sight value $N_{\rm H} = 5\times10^{20}$ cm$^{-2}$ for all the 
three spectra.

The stacked spectrum of the hard state is well fit by a single power-law model with photon 
index $\Gamma \approx 1.6$ (\autoref{fit}), similar to the typical values found in the hard state
of LMXBs. The stacked spectrum in the soft state is fit by a disk-blackbody 
model plus a power-law component that accounts for significant residual emission above 2 keV. 
The disk contributes $\approx$85\% of the 0.3--10 keV flux, similar to the situation in the 
disk-dominated soft state of BH LMXBs. The peak color temperature $T_{\rm in}$ in the soft state 
is $\approx$0.2 keV (\autoref{fit}), which is much less than the typical value in luminous BH LMXBs 
($T_{\rm in}\approx1$ keV). The corresponding inner-disk radius $R_{\rm in} \sim 50,000$--
$100,000$ km for a plausible range (10$^{\circ}$--80$^{\circ}$) of disk inclination angles, at a distance of 95 Mpc.  The low color temperature and large radius of the thermal 
emitter are consistent with a disk around an IMBH \citep{Godet2009,Davis2011,Servillat2011}. However, 
the thermal X-ray emission components with cool temperatures and large radii are also seen in 
ultraluminous supersoft sources where the emission is probably coming from the photosphere of 
a thick outflow rather than from a disk \citep{Poutanen2007, Shen2015}, and therefore do not 
require an IMBH as the accreting object \citep[\textit{e.g.} the supersoft ULX in M\,101 is probably 
powered by a stellar-mass BH,][]{Liu2013}. Finally, the stacked spectra of HLX-1 in the 
transitional state are well fit by a disk-blackbody ($T_{\rm in} \approx 0.16$ keV) plus 
power-law ($\Gamma\approx2.45$) model (\autoref{fit}), where the disk emission 
accounts for $\approx$52\% of the 0.3--10 keV flux. The disk normalization is similar to the soft state (\autoref{fit}), which is consistent with a disk extending down to the inner-most stable circular orbit (ISCO).

\subsection{State Transition Luminosity}
\label{sec2.2}
\begin{figure*}  
\centering    
\includegraphics[width=\linewidth]{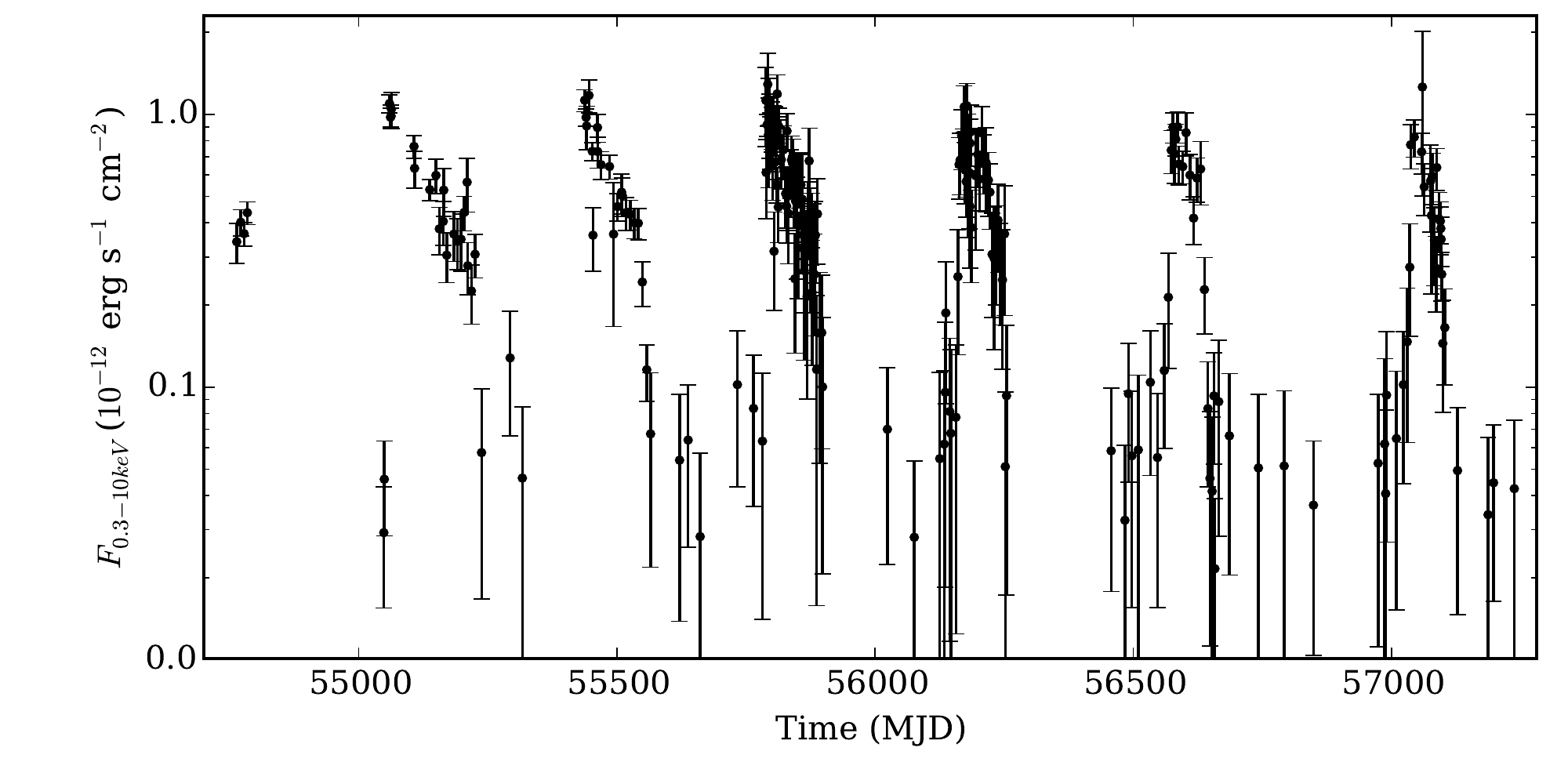}
\caption{Complete {\it Swift}/XRT light curve 
(unabsorbed flux in unit of erg s$^{-1}$ cm
$^{-2}$), binned to individual observations.}
\vspace{0.5cm}
\label{lc}
\end{figure*}

According to our definition of the three spectral states, the mean XRT count rates are $
\approx$0.022 ct s$^{-1}$, $\approx$0.005 ct s$^{-1}$ and $\approx$0.012 ct s$^{-1}$ for the 
soft, hard and transitional states, respectively. Based on our spectral fitting results 
(\autoref{fit}), the unabsorbed 0.3--10 keV fluxes corresponding to an XRT count rate of 
1 ct s$^{-1}$ are $3.11\times10^{-11}$ erg s$^{-1}$ cm$^{-2}$ in the soft state, 
$2.60\times10^{-11}$ erg s$^{-1}$ cm$^{-2}$ in the hard state, and $3.03\times10^{-11}$ 
erg s$^{-1}$ cm$^{-2}$ in the transitional state. For each observation, we used the hardness 
ratio to identify the spectral state, then converted the count rate to an unabsorbed flux by using 
the appropriate conversion factor. The {\it Swift}/XRT light curve in units of erg s$^{-1}$ cm
$^{-2}$ is shown in \autoref{lc}.

The {\it Swift} observations have so far covered seven outbursts of HLX-1, plus only a few 
data points for an outburst in 2008 (\autoref{lc}); we exclude those 2008 
data points in the rest of this paper. We determined the luminosity $L_\mathrm{hard-to-soft}$ at 
the transition from hard to soft state
during the outburst rise, and the luminosity $L_\mathrm{soft-to-hard}$
at the reverse transition during the decay phase, using the same
method previously applied by \citet{Yu2009} to Galactic LMXBTs. In
that method, we used the luminosity of the last hard state data point
as the hard-to-soft state transition luminosity
\citep{Yu2009,Tang2011}. However, it is much harder to catch the
actual transition in the HLX-1 outbursts because the rise time is
short compared with the {\it Swift} sampling frequency. Therefore, we note
that we might be underestimating $L_\mathrm{hard-to-soft}$ with our
method. The {\it Swift} observations do not cover the hard-to-soft
state transition in the 2009 and 2010 outbursts; for the other four
outbursts, we list $L_\mathrm{hard-to-soft}$ in \autoref{trans}. The
first observation with HR $> 0.6$ during the decay phase was
identified as the first hard state observation; we used the luminosity
of this observation as the soft-to-hard state transition luminosity
$L_\mathrm{soft-to-hard}$ \citep{Yu2009}. We list the
$L_\mathrm{soft-to-hard}$ for six outbursts in \autoref{trans}.

\begin{figure}  
\centering    
\includegraphics[width=\linewidth]{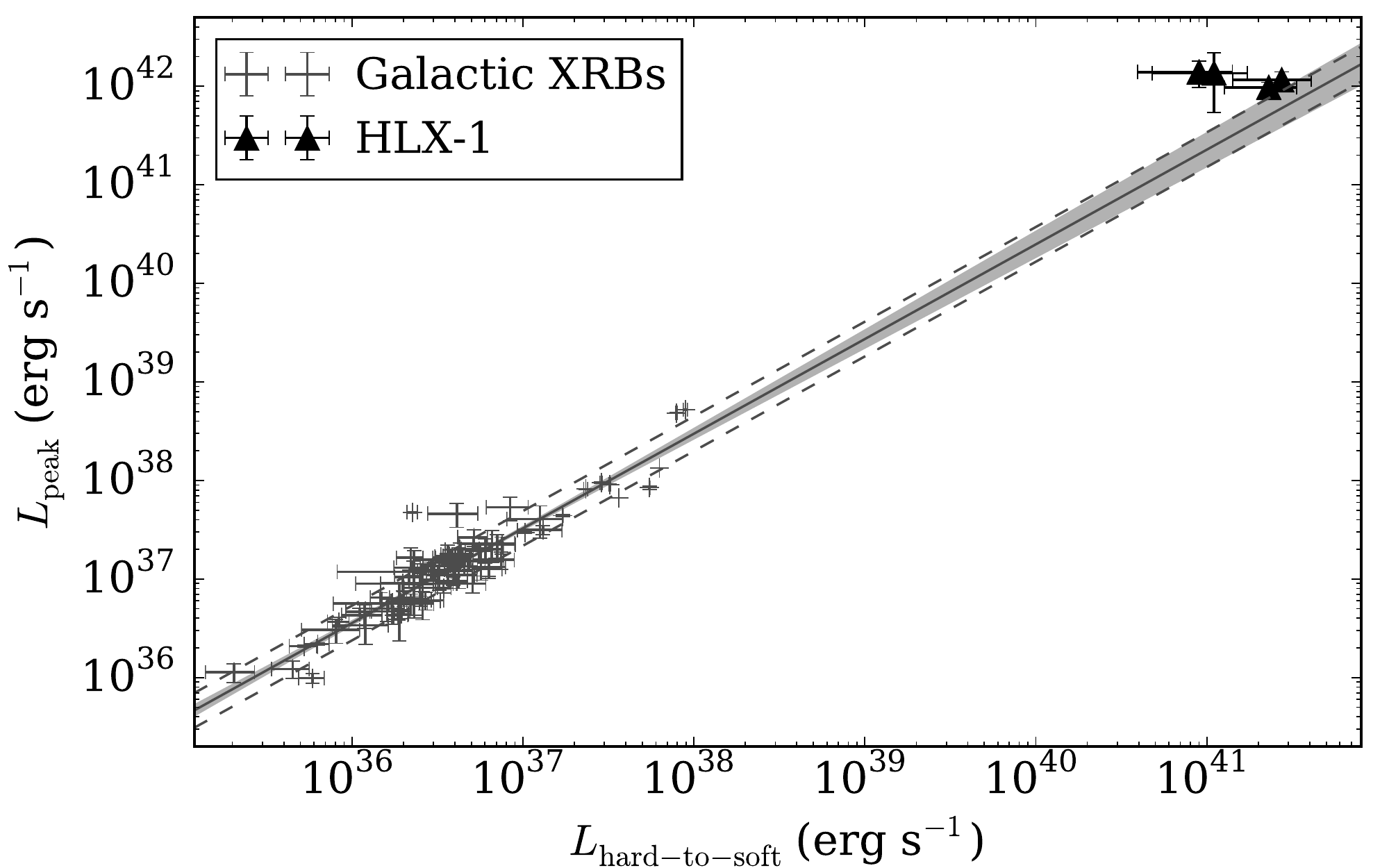}
\caption{Correlation between $L_\mathrm{hard-to-soft}$ and $L_\mathrm{peak}$ in Galactic 
LMXBTs, and the location of HLX-1 outbursts (plotted as triangles) in the same parameter space. 
The solid line is a linear regression of the form $\log L_\mathrm{peak} = A+B\times\log L_
\mathrm{hard-to-soft}$, fit to the Galactic LMXBT sample using the method of 
\citet{Kelly2007}. The dashed lines show the intrinsic scatter and the shaded region shows the 
1$\sigma$ confidence interval of the regression line. HLX-1 is roughly consistent with the 
correlation.}
\vspace{0.5cm}
\label{h2s}
\end{figure}

We had already found a correlation between the hard-to-soft state
transition luminosity and the peak luminosity in Galactic XRBs
\citep{Yu2004,Yu2007,Yu2009,Tang2011}. We fit the correlation in Galactic XRBs with 
a function $\log L_\mathrm{peak} = A+B\times\log L_\mathrm{hard-to-soft}$ by using the 
Bayesian approach in Kelly (2007), and obtained a best-fitting slope of B = 0.96$\pm$0.04, 
which means that
$L_\mathrm{hard-to-soft}$ is linearly proportional to $L_\mathrm{peak}$. 
Now we have added the values of $L_\mathrm{hard-to-soft}$ and $L_\mathrm{peak}$ for 
the four HLX-1 outbursts (\autoref{h2s}). HLX-1 roughly follows this correlation we found in Galactic XRBs, 
which confirms the prediction that the correlation extends to ULXs \citep{Yu2009}.

\capstartfalse
\begin{deluxetable}{ccc}
\centering
\tablecaption{State Transition Luminosity}
\tablewidth{0pt}
\tablehead{
\colhead{Outburst} &\colhead{$L_\mathrm{hard-to-soft}$}      & \colhead{$L_\mathrm{soft-
to-hard}$}   \\
\colhead{} &\colhead{(10$^{41}$ erg s$^{-1}$)}      & \colhead{(10$^{41}$ erg s$^{-1}$)}
 }
\startdata
2009 & \nodata              &$0.62\pm0.44$   \\
2010 & \nodata              & $1.25\pm0.29$  \\
2011 &$0.90\pm0.51$   &$2.32\pm1.35$ \\
2012 &$2.74\pm1.32$   &  $0.55\pm0.48$         \\
2013 &$2.30\pm1.04$   &  $0.90\pm0.44$ \\
2015 &$1.10\pm0.62$   &  $1.78\pm0.69$ \\
\enddata
\label{trans}
\end{deluxetable}
\capstarttrue

\subsection{Outburst Parameters}
\label{sec2.3}
 \label{op}
 \begin{figure*}  
\centering    
\includegraphics[width=\linewidth]{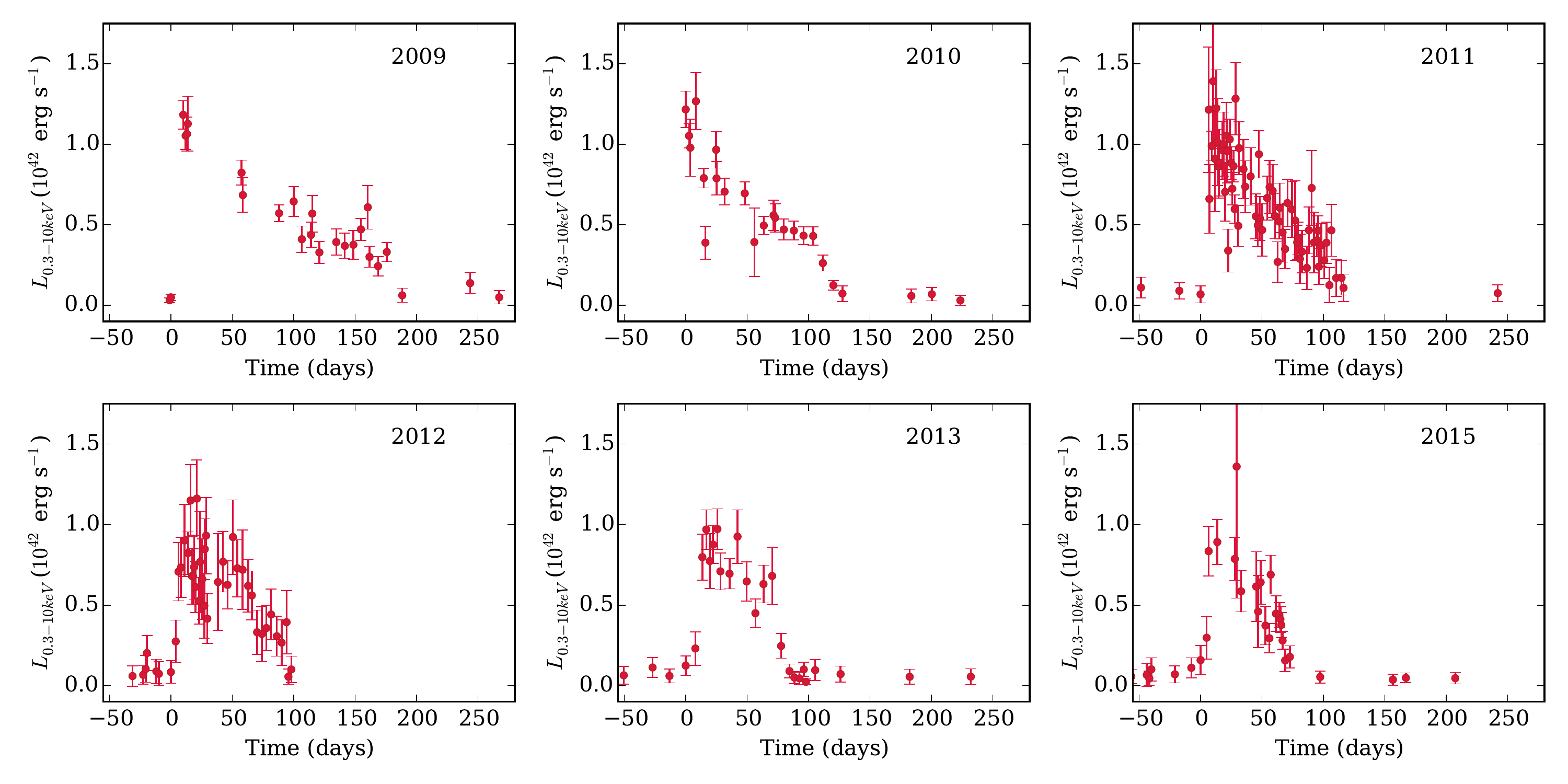}
\caption{{\it Swift}/XRT light curves of each outburst (excluding the
  few sparse data points in 2008). In each panel, Time $= 0$ corresponds to the time of $L_\mathrm{rise,10\%}$.}
  \vspace{0.5cm}
\label{olc}
\end{figure*}

\begin{figure}  
\centering    
\includegraphics[width=\linewidth]{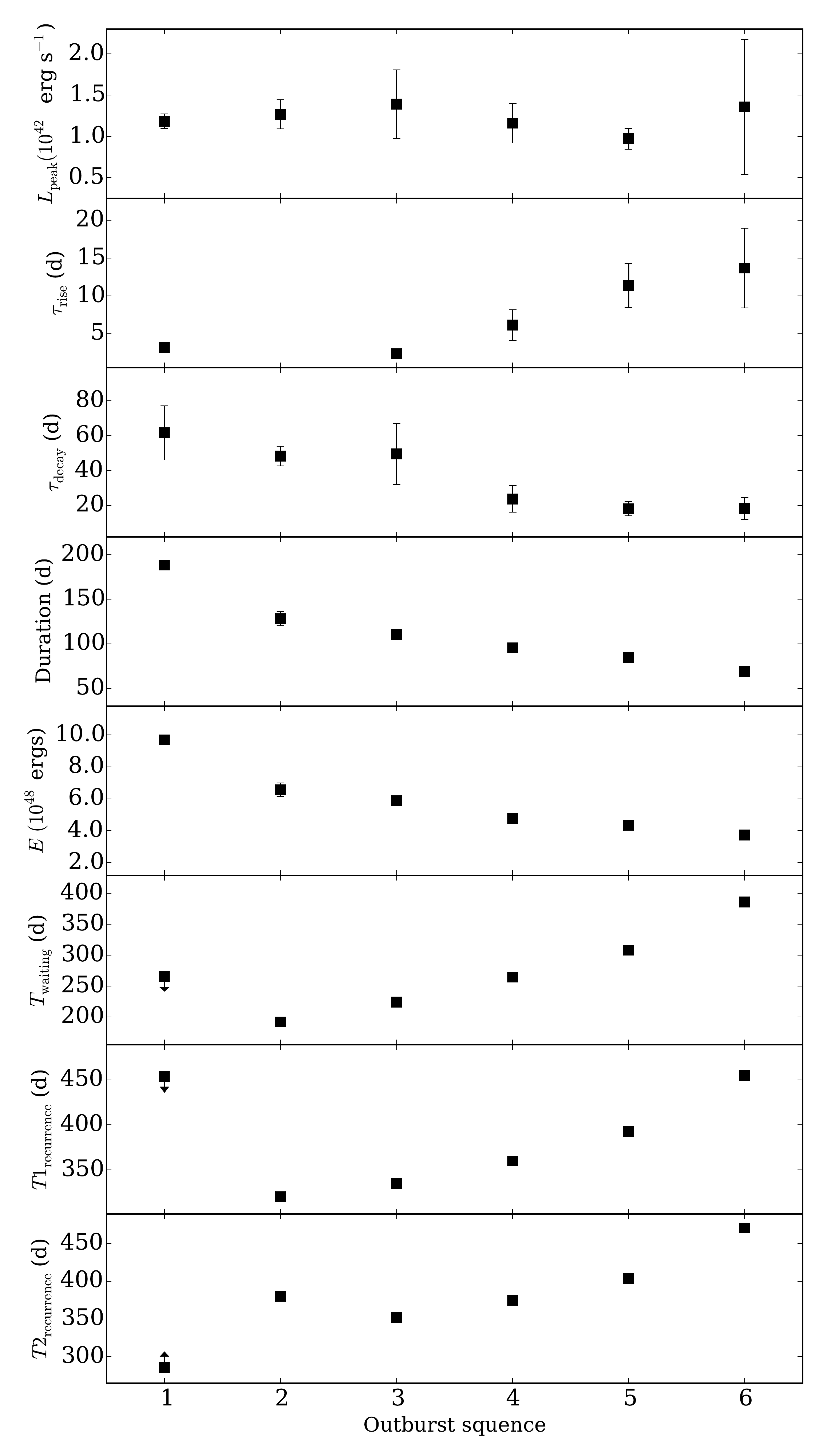}
\caption{Evolution of the main outburst parameters defined in \autoref{sec2.3}. The total radiated energy $E$, the outburst duration, and the
  $e$-folding decay timescale $\tau_\mathrm{decay}$ decreased and the
  waiting time, recurrence time, and $e$-folding rise timescale $\tau_\mathrm{rise}$
  increased. The first outburst is the one in 2009. $T_\mathrm{waiting}$ is the waiting time.
  $T1_\mathrm{recurrence}$ is the recurrence time defined as the interval between 
  $L_\mathrm{decay,10\%}$ of an
outburst and $L_\mathrm{decay,10\%}$ of the previous outburst. $T2_\mathrm{recurrence} $
is the recurrence time defined as the interval between $L_\mathrm{rise,10\%}$ of an
outburst and $L_\mathrm{rise,10\%}$ of the previous outburst} 
  \vspace{0.5cm}
\label{opars}
\end{figure}

\capstartfalse
\begin{deluxetable*}{ccccccccccc}
\tablecaption{Outburst parameters}
\tabletypesize{\tiny}
\centering
\tablehead{\colhead{Peak Time} & \colhead{ $L_\mathrm{peak}$} & 
\colhead{Duration} & \colhead{$T_\mathrm{waiting}$ \tablenotemark{a}} & \colhead{$T1_{\rm recurrence}$ \tablenotemark{b}}
&\colhead{$T2_{\rm recurrence}$ \tablenotemark{c}} & \colhead{$E$ \tablenotemark{d}} & \colhead{$
\mathrm{\tau_\mathrm{rise}}$} & \colhead{$\mathrm{\tau_\mathrm{decay}}$} & \colhead{$
\dot{L}_\mathrm{rise}$} & \colhead{$\dot{L}_\mathrm{decay}$} \\
\colhead{} & \colhead{($\times10^{42}$)} & \colhead{} & \colhead{} & \colhead{}& \colhead{}& \colhead{($
\times10^{48}$)} & \colhead{} & \colhead{} & \colhead{($\times10^{40}$)} & \colhead{($
\times10^{40}$)}\\
\colhead{(MJD)} & \colhead{($\mathrm{erg~s^{-1}}$)} & \colhead{(days)} & 
\colhead{(days)} & \colhead{(days)} & \colhead{(days)} & \colhead{($\mathrm{erg}$)} 
& \colhead{(days)} & \colhead{(days)} 
&\colhead{($\mathrm{erg~s^{-1}~day^{-1}}$)} & \colhead{($\mathrm{erg~s^{-1}~day^{-1}}$)} 
}
\startdata
55060 & 1.2$\pm$0.1 & 188 & $<$265 & $<$454 &$>$286& 9.7$\pm$0.1 & 3$\pm$0 & 62$\pm$16 & 11.3$\pm$0.9 & 0.6$\pm$0.1 \\
55446 & 1.3$\pm$0.2 & 128$\pm8$ & 192$\pm8$ & 320 &380 & 6.6$\pm$0.4 &\nodata & 48$\pm$6 &\nodata & 1.0$\pm$0.2 \\
55792 & 1.4$\pm$0.4 & 110 & 224 & 334 & 352 & 5.9$\pm$0.2 & 2$\pm$1 & 50$\pm$17 & 17.2$\pm$5.9 & 1.1$\pm$0.3 \\
56178 & 1.2$\pm$0.2 & 96 & 264 & 360 &375 & 4.8$\pm$0.2 & 6$\pm$2 & 25$\pm$8 & 7.0$\pm$1.5 & 1.4$\pm$0.3 \\
56586 & 1.0$\pm$0.1 & 84 & 308 & 392 &404 & 4.3$\pm$0.1 & 11$\pm$3 & 18$\pm$4 & 3.4$\pm$0.6 & 2.0$\pm$0.4 \\
57060 & 1.4$\pm$0.8 & 69 & 386 &455 &470 & 3.7$\pm$0.1 & 14$\pm$5 & 18$\pm$6 & 4.1$\pm$2.8 & 3.0$\pm$2.1 \\
\enddata
\label{pars}

\tablenotetext{a}{  Waiting time }
\tablenotetext{b}{  Recurrence time defined as the interval between $L_\mathrm{decay,10\%}$ 
of an outburst and $L_\mathrm{decay,10\%}$ of the previous outburst}
\tablenotetext{c}{  Recurrence time defined as the interval between $L_\mathrm{rise,10\%}$ of an
outburst and $L_\mathrm{rise,10\%}$ of the previous outburst}
\tablenotetext{d}{  Total radiated energy per outburst }

\end{deluxetable*}
\capstarttrue

The shape of the X-ray light curves for the six HLX-1 outbursts with
good {\it Swift} coverage are qualitatively similar. The 2009 and 2011
outbursts show the clearest FRED profiles (\autoref{olc}), which is
typical of Galactic LMXBTs outbursts \citep{Chen1997}. We only have
flux upper limits for the early rise phase of the 2010 outburst, but
they are also consistent with a FRED profile (\citealt{Soria2013a},
Fig.~1). The 2012, 2013, and 2015 outbursts are roughly consistent with FRED
profiles, but also show secondary peaks or reflarings during the
decline (\autoref{olc}).

We measured the main outburst parameters for the six outbursts
(\autoref{pars}, \autoref{opars}), including peak luminosity,
$e$-folding rise and decay timescales, rate of change of the
luminosity during the rise and decay, outburst duration, and total
radiated energy, according to the definitions in \citet{Yan2015}. The
peak luminosity $L_\mathrm{peak}$ is defined as the maximum luminosity
in each outburst. We defined $L_\mathrm{rise,10\%}$,
$L_\mathrm{decay,10\%}$, $L_\mathrm{rise,90\%}$ and
$L_\mathrm{decay,90\%}$ as the measured data points closest to $10\%$
and $90\%$ of $L_\mathrm{peak}$ during the rise and decay phases, 
respectively. The $e$-folding rise timescale is
defined as $\tau_\mathrm{rise} \equiv {\Delta t}/{\ln A}$, where
${\Delta t}$ is the time interval between the data points corresponding
to $L_\mathrm{rise,10\%}$ and $L_\mathrm{rise,90\%}$, and $A$ is the
ratio of $L_\mathrm{rise,90\%}/L_\mathrm{rise,10\%}$. The $e$-folding
decay timescale $\tau_\mathrm{decay}$ is defined in the same way, but
for $L_\mathrm{decay,10\%}$ and $L_\mathrm{decay,90\%}$. The 
rate of change of luminosity $\dot{L}_\mathrm{rise}$ and
$\dot{L}_\mathrm{decay}$ are defined as
$(L_\mathrm{rise,90\%}-L_\mathrm{rise,10\%})/\Delta t$ for the rise
phase and as $(L_\mathrm{decay,90\%}-L_\mathrm{decay,10\%})/\Delta t$
for the decay phase, respectively. 
The outburst duration is defined as the time interval between
$L_\mathrm{rise,10\%}$ and $L_\mathrm{decay,10\%}$.  We have two definitions of the
recurrence time, one is the interval between $L_\mathrm{decay,10\%}$ of an
outburst and $L_\mathrm{decay,10\%}$ of the previous outburst (abbreviated as $T1_\mathrm{recurrence}$); 
the other is the interval between $L_\mathrm{rise,10\%}$ of an
outburst and $L_\mathrm{rise,10\%}$ of the previous outburst (abbreviated as $T2_\mathrm{recurrence}$). The
total radiated energy $E$ is defined as the integrated luminosity from
the time of $L_\mathrm{rise,10\%}$ to the time of
$L_\mathrm{decay,10\%}$ (for this integration, the data gaps due to
sparse coverage were filled with linear interpolations). Finally, we
determined the waiting time for each outburst, defined as the time
interval between $L_\mathrm{rise,10\%}$ of an outburst and
$L_\mathrm{decay,10\%}$ of the previous outburst. For the outburst in
2009, we only have upper limits to the waiting time and $T1_\mathrm{recurrence}$ and lower limits to the $T2_\mathrm{recurrence}$,  because the
data coverage of the 2008 outburst is too limited to permit the
identification of  the times of $L_\mathrm{rise,10\%}$ and $L_\mathrm{decay,10\%}$.

The uncertainties in the outburst durations,  waiting times, and recurrence times are
mainly due to the {\it Swift}/XRT observation sampling frequency. The
typical interval of observations in different outbursts is in the
range of 1--14 days. For example, for the 2010 outburst we are not able
to identify $L_\mathrm{rise,90\%}$ (\autoref{olc}); therefore, we
cannot calculate the $\tau_\mathrm{rise}$ and
$\dot{L}_\mathrm{rise}$. Instead, we use the earliest significant
detection (\autoref{olc}) as $L_\mathrm{rise,10\%}$ to estimate lower
limits to the duration, total radiated energy, and waiting time of the
2010 outburst. The observation before the earliest significant
detection is only an upper limit: by taking that value as
$L_\mathrm{rise,10\%}$, we estimated upper limits to the duration,
total radiated energy, and waiting time (\autoref{pars}).

Our main findings are the following.
\begin{enumerate}
\item There is no clear trend for the peak luminosity with
  outburst sequence, probably because it is well constrained in
  the range of 1.0--1.4 $\times10^{42}\mathrm{~erg~s^{-1}}$
  (\autoref{opars}, \autoref{pars}). The value and uncertainty of the peak luminosity of the 2015 outburst is quite large, which could be due to the short-term variability, since the exposure time of this observation is short. \citet{Godet2013, Godet2014}
  argued that the peak count rate decreases with outburst sequence,
  but we did not find such a trend in the peak luminosity. There may be
  two reasons for this discrepant conclusion. Firstly, our light curve
  was rebinned by observation, while \citet{Godet2013, Godet2014}
  rebinned theirs by photon count (20 counts per bin): as a result, the
  peak values are slightly different. Secondly, our light curve is in
  flux rather than count rate units: the conversion factor is
  different depending on whether the peak of the outburst was reached
  during the transitional or soft state.

\item The $e$-folding rise timescale increases at least since the 2011 outburst.

\item The $e$-folding decay timescale decreases, which confirms the result in \citet{Miller2014}.

\item The outburst duration also decreases along the outburst sequence
  , which confirms previous results in \citet{Godet2013,Godet2014}, from $\approx$188
  days in the 2009 outburst to $\approx$69 days in the 2015 outburst
  (\autoref{opars}, \autoref{pars}).
  
\item The total radiated energy decreases along the outburst sequence, 
  which confirms the result in \citet{Miller2014}, from $E = 9.7\pm0.1 \times
  10^{48}$ erg in the 2009 outburst to $E = 3.7\pm0.1 \times
  10^{48}$ erg in the 2015 outburst (\autoref{opars} and
  \autoref{pars}). This corresponds to a mass transfer of $\approx$$5
  \times 10^{-5} M_{\odot}$ in the 2009 outburst, down to $\approx$$2
  \times 10^{-5} M_{\odot}$ in the 2015 outburst (assuming a standard 
  efficiency $\approx$0.1).
  
\item The waiting time increases along the outburst sequence at least
  since the 2010 outburst, from $\approx$192 days for the 2010
  outburst to $\approx$386
  days for the 2015 outburst (\autoref{opars}, \autoref{pars}).

\item  The recurrence time increases along the outburst sequence, from $\approx$ 320 days of the 2010 outburst to $\approx$ 455 days of the 2015 outburst for $T1_\mathrm{recurrence}$, and from $\approx$ 352 days of the 2011 outburst to $\approx$ 470 days of the 2015 outburst for $T2_\mathrm{recurrence}$ (\autoref{opars}, \autoref{pars}). The increase of the recurrence time creates serious problems for its original and simplest interpretation as the binary period. Therefore, \citet{Godet2014} tried to resolve this problem by considering a binary system with extreme parameters, and they predicted an increase of the binary period in such system, which is consistent with the increase of recurrence time \citep{Godet2015}.

\item The duty cycle (defined as the fraction of time spent in
outbursts) is $\approx$0.3.
\end{enumerate}

\subsection{Comparison of Outburst Properties with Bright Galactic LMXBTs}
\label{sec2.4}
\begin{figure} 
\centering   
\includegraphics[width=\linewidth]{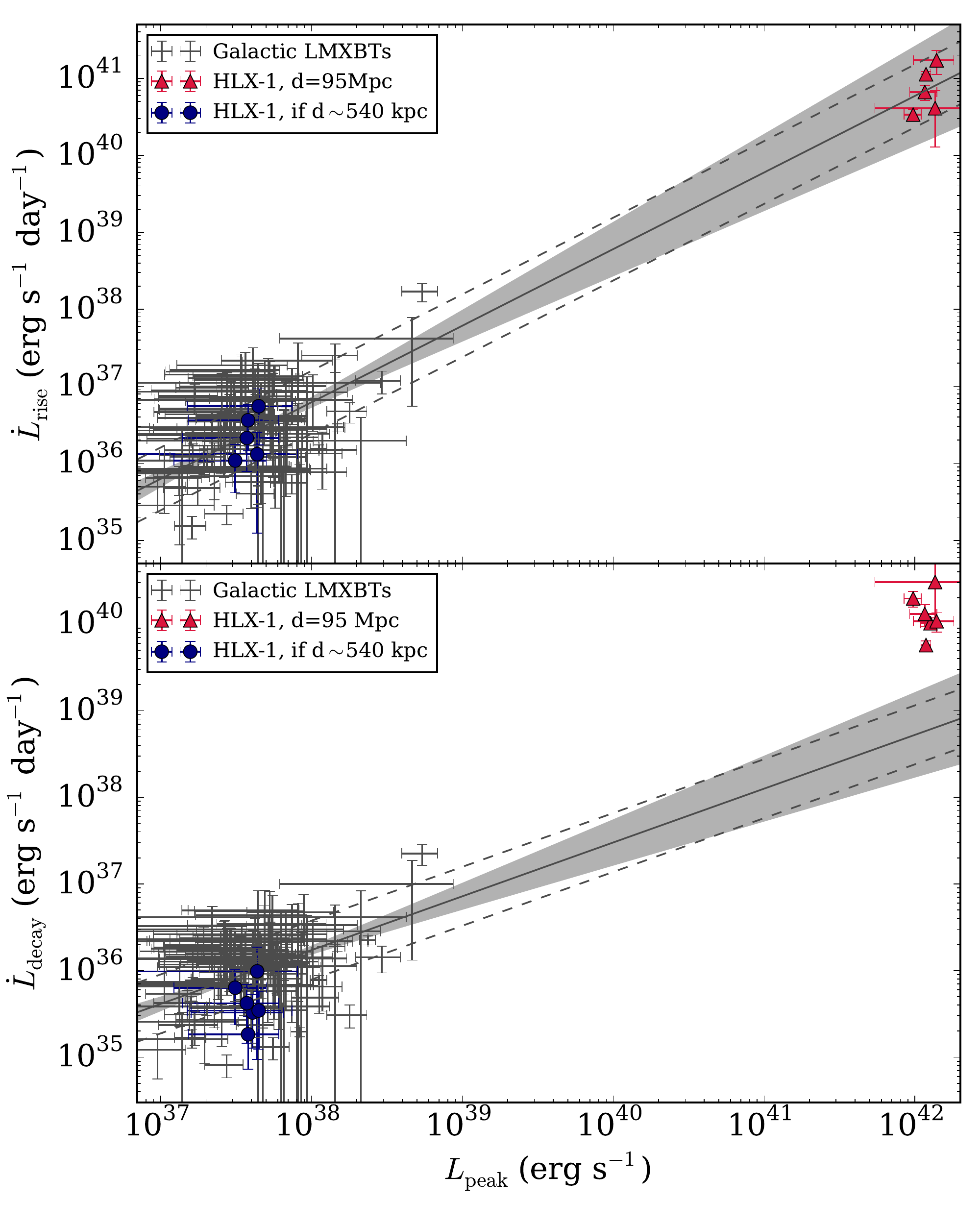}
\caption{Relations between $\dot{L}$ and
   $L_\mathrm{peak}$ during outburst rise (top panel)
  and decay (bottom panel) in Galactic LMXBTs and HLX-1. The solid
  line shows the best-fitting linear correlation for the Galactic
  sources, with their intrinsic scatter (dashed lines) and 1$\sigma$
  confidence intervals (shaded regions). HLX-1 follows the correlation
  between $\dot{L}_\mathrm{rise}$ and $L_\mathrm{peak}$, but not the
  one between $\dot{L}_\mathrm{decay}$ and $L_\mathrm{peak}$. HLX-1 would
  have to be located at $\approx$540 kpc to satisfy the both correlations, 
  which is inconsistent with its optical redshift (see details in 
  \autoref{sec3.3}). 
  }
  \vspace{0.5cm}
\label{ldot}
\end{figure} 

\begin{figure}  
\centering    
\includegraphics[width=\linewidth]{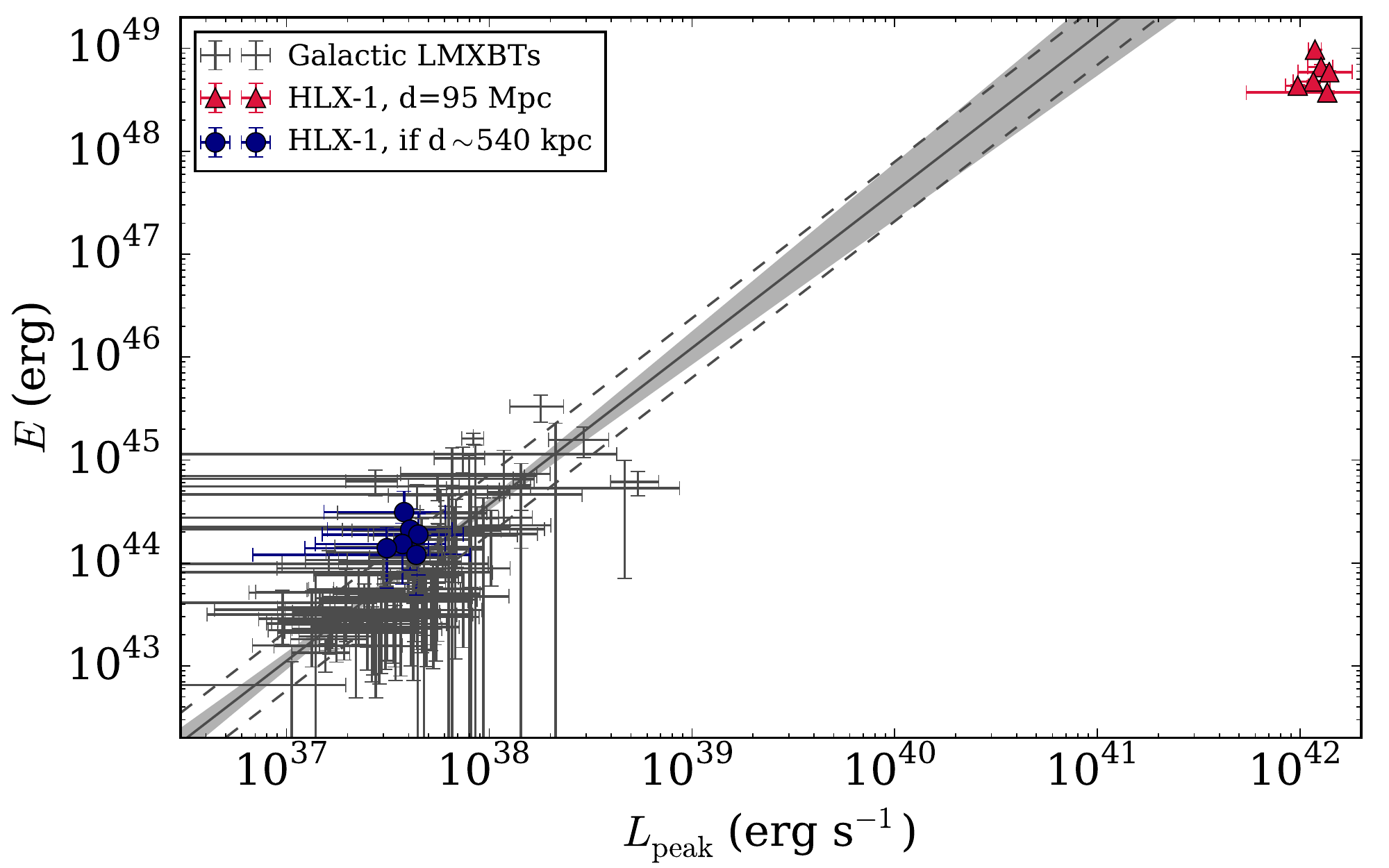}
\caption{Relation between $E$ and $L_\mathrm{peak}$ in Galactic LMXBTs and HLX-1.  The
  solid line shows the best-fitting correlation for the Galactic
  sources, with their intrinsic scatter (dashed lines) and 1$\sigma$
  confidence intervals (shaded regions). HLX-1 does not follow this
  correlation. As for the other cases shown in Figure 8, HLX-1 will follow this 
  correlation if it is located at a distance of $\approx$540 kpc (see details in 
  \autoref{sec3.3}), which is inconsistent with its optical redshift. }
  \vspace{0.5cm}
\label{epeak}
\end{figure} 

\begin{figure}  
\centering    
\includegraphics[width=\linewidth]{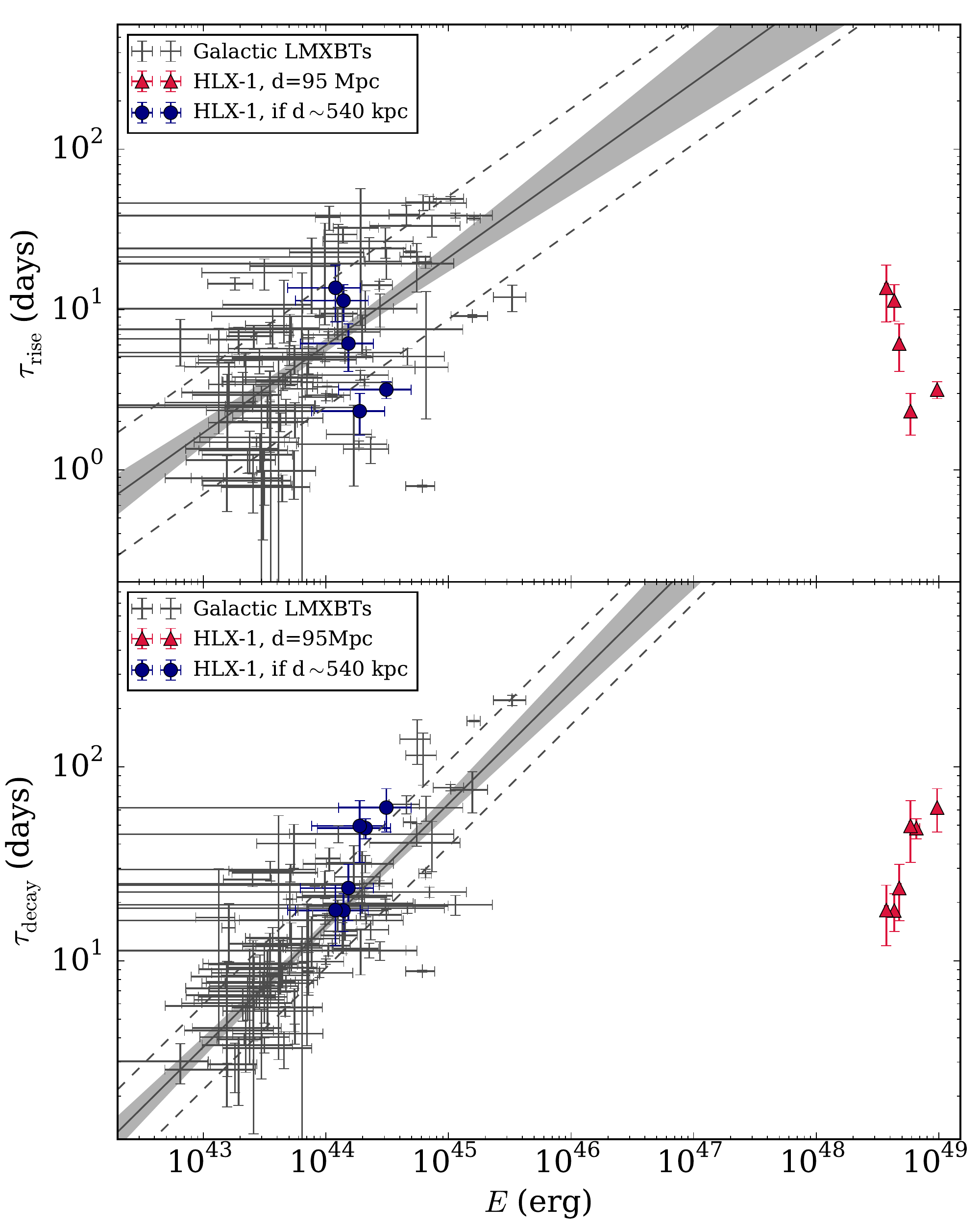}
\caption{Relation between $e$-folding rise (top panel) and decay
  (bottom panel) timescales $\tau$ and total radiated energy $E$, for
  Galactic LMXBTs and HLX-1. Solid and dashed lines, and shaded
  regions, are defined as in Figures 8 and 9. HLX-1 is several orders
  of magnitude off the correlation. In this case, too, it would
  satisfy the LMXBT correlation if it were at $\approx$540 kpc (see details in 
  \autoref{sec3.3}).}
  \vspace{0.5cm}
\label{etau}
\end{figure} 

The comparison between the outburst properties of HLX-1 and those we
determined previously for bright Galactic LMXBTs \citep{Yan2015}
can be summarized as follows. 

\begin{enumerate}

\item The average peak luminosity of HLX-1 ($\approx1.2 \times
  10^{42}$ erg s$^{-1}$) is more than four orders of magnitude larger than the
  average peak luminosity of bright Galactic LMXBTs
  ( $\approx7 \times 10^{37}$ erg s$^{-1}$ and $\approx 4\times 10^{37}$ erg s$^{-1}$ for BH and NS LMXBTs, respectively).

\item The average total radiated energy per outburst of HLX-1
  ($\approx5.8\times10^{48}$ erg) is about five orders of magnitude larger
  than those measured in bright Galactic LMXBTs
  ( $\approx2\times10^{44}$ erg and $\approx 5\times 10^{43}$ erg for BH and NS LMXBTs, respectively).

\item The average outburst duration of HLX-1 ($\approx$ 113 days) is
  about a factor of 3 longer than that of  the NS LMXBTs
  outbursts ($\approx$ 39 days), and is a little bit longer than the
  average outburst duration of BH LMXBTs
  ($\approx$ 88 days), but still within the $1\sigma$ range of BH LMXBTs.

\item The average $e$-folding rise timescale of HLX-1 ($\approx$7 days) is about
  two times larger than the Galactic NS LMXBTs ($\approx$4 days), and is similar to the Galactic BH LMXBTs ($\approx$8 days).

\item The average $e$-folding decay timescale of HLX-1 ($\approx$37
  days) is a few times longer than for Galactic NS LMXBTs (average of
  $\approx$11 days) and BH LMXBTs (average of $\approx$26
  days),   but still within the $1\sigma$ range of values measured for BH LMXBTs.

\end{enumerate}

Positive correlations have been found between the 
rate of change of luminosity and the peak luminosity in both the rise and decay phases, for Galactic
LMXBTs  \citep{Yan2015}. Using a Bayesian linear regression \citep{Kelly2007}, we had
found that $\log \dot{L}_{\mathrm{rise}} = A+B\times\log
L_\mathrm{peak}$ with $B=0.99\pm0.20$, and $\log
\dot{L}_{\mathrm{decay}} = A'+B'\times\log L_\mathrm{peak}$ with
$B'=0.62\pm0.16$. We extrapolated those empirical relations to see
whether HLX-1 is also consistent with them (\autoref{ldot}). We found
that HLX-1 follows the correlation between $\dot{L}_\mathrm{rise}$ and
$L_\mathrm{peak}$, but not the one between $\dot{L}_\mathrm{decay}$
and $L_\mathrm{peak}$ \citep[{\it i.e.}, its decay timescale is too short to
be a scaled-up LMXBT; see also][]{Lasota2015}. Other correlations found in Galactic LMXBTs are
those between the total radiated energy and peak luminosity, and those
between the $e$-folding rise/decay timescales and total radiated energy;
none of those relations is satisfied by HLX-1
(\autoref{epeak} and \autoref{etau}). (Probably only by coincidence, HLX-1 would satisfy
those Galactic LMXBTs correlations if it were located at a distance of
$\sim$540 kpc, which is inconsistent with its optical line redshift
and is implausible based on other arguments, see \autoref{sec3.3}.).

\subsection{Comparison with BH LMXBT H1743$-$322}
\label{sec2.5}
\begin{figure*}  
\centering    
\includegraphics[width=\linewidth]{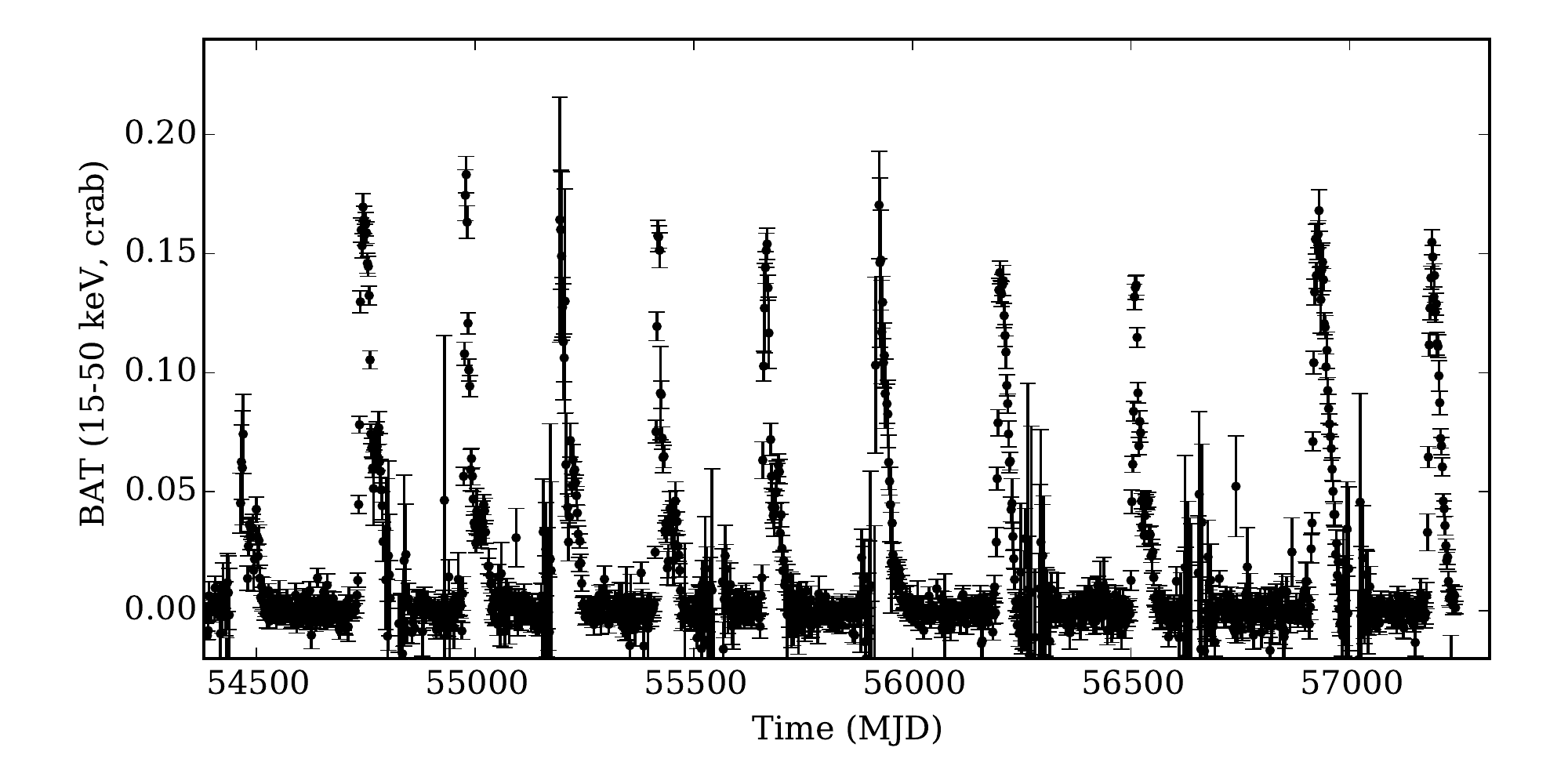}
\caption{Long-term X-ray light curve of H1743$-$322 measured by {\it
    Swift}/BAT.  It shows repeated outbursts nearly
  each year after MJD 54700, similar to the outburst sequence of
  HLX-1.}
  \vspace{0.5cm}
\label{blc}
\end{figure*} 

\begin{figure}  
\centering    
\includegraphics[width=\linewidth]{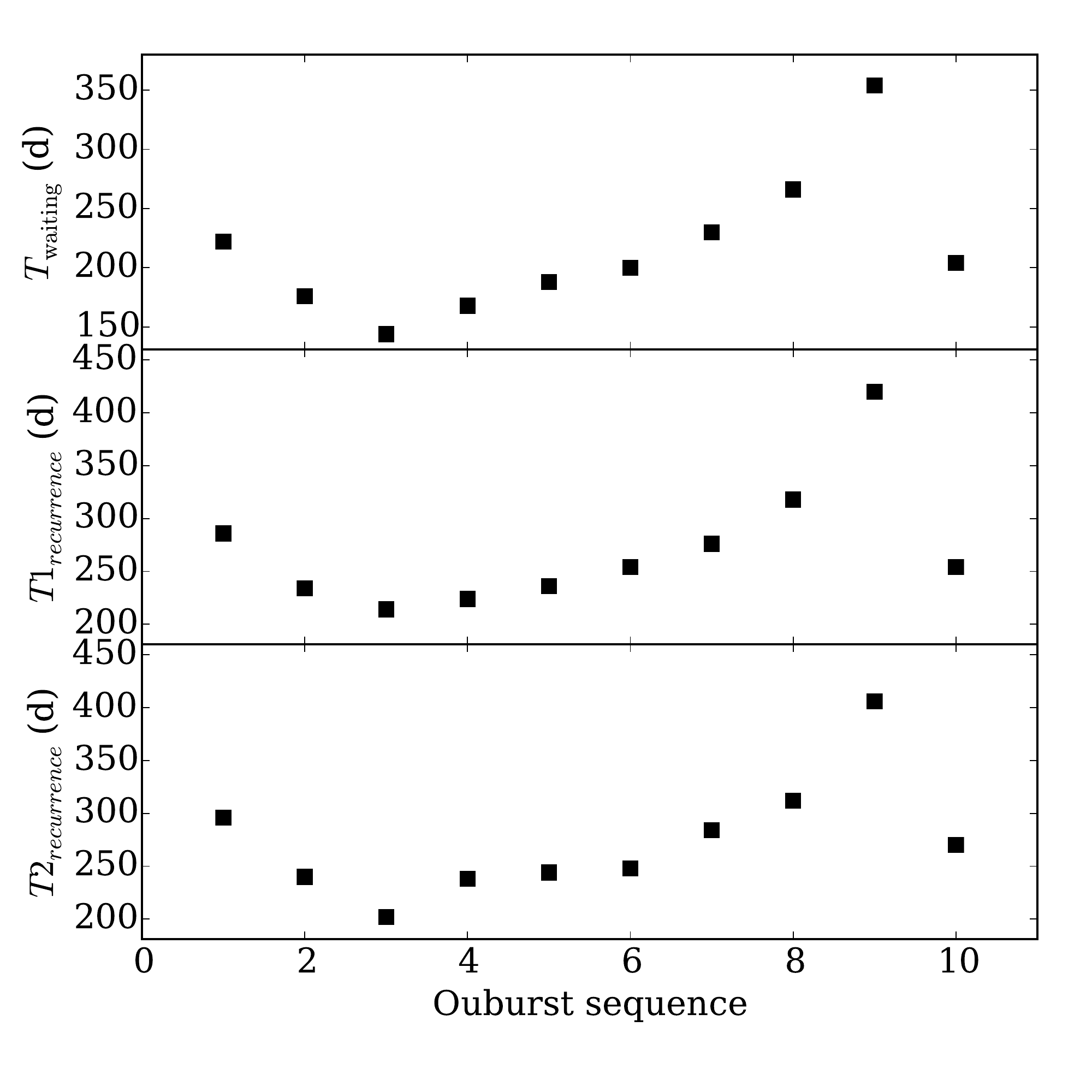}
\caption{Evolutions of the waiting time and recurrence time for the most recent outbursts
  of H1743$-$322.  The zero outburst corresponds to the outburst peaking on about MJD 54470. 
  The $T_\mathrm{waiting}$, $T1_\mathrm{recurrence}$ and $T2_\mathrm{recurrence} $
are defined in the same way to HLX-1.}
\vspace{0.5cm}
\label{hwt}
\end{figure} 

In addition to comparing the outburst properties of
HLX-1 with the average outburst properties of bright
Galactic LMXBTs  (\autoref{sec2.4}), we compare HLX-1 with some
individual systems which undergo
regular outbursts. For example, the BH
LMXBT H1743$-$322 has been in outburst nearly
once a year for the last 10 years (\autoref{blc}), as shown
by the {\it Swift}/BAT light curve \citep{Krimm2013}.
We used that light curve to quantify the outburst properties
of H1743$-$322, and investigate its similarities with HLX-1.
We only considered those outbursts with a peak count rate larger
than 0.1 crab at a $2\sigma$ significance level; this means
all outbursts after MJD 54700 ({\it i.e.}, after 2008 August).
We used the same method described in \autoref{sec2.3}
to measure the outburst parameters.
The outbursts occur quite regularly with
an average recurrence time of $274 \pm 56$ days, which is very similar
to that of HLX-1 (\autoref{lc}).
We found that the waiting time and recurrence time first decrease and then
increase (see \autoref{hwt}), possibly consistent with the behavior of HLX-1 
(bottom panels of \autoref{opars}). However, the waiting time and recurrence time 
suddenly decrease to the average level in the latest 2015 outburst \autoref{hwt}.
 Instead, we did not find any clear
trend in the outburst duration, fluence, peak flux, and $e$-folding rise
and decay timescales versus outburst sequence. 

We repeated the outburst analysis for H1743$-$322 using {\it RXTE}/ASM
and MAXI data \citep[2--10 keV;][]{Matsuoka2009}. We obtained similar values
for the waiting and recurrence times; this shows that such parameters are not
significantly affected by the choice of energy band. We do not have
enough evidence at this stage to test whether or not the possible similarity
in the outburst waiting and recurrence times between HLX-1 and H1743$-$322 points
to the same physical mechanism for outburst triggering and evolution.

For completeness, we also performed a similar comparison
between the BH LMXBT GX~339$-$4,  4U 1630$-472$, the NS LMXBT Aql X-1 (using
$RXTE$/ASM+MAXI data), and HLX-1. These three LMXBTs also show frequent
outbursts. However, for GX~339$-$4 and 4U 1630$-472$, we do not find any clear trends
when using data from their past 5 outbursts and 10 outbursts, nor we find that for Aql
X-1 when using data from its past 14 outbursts.

\section{DISCUSSION}
\subsection{BH Mass Estimation Based on $L_{soft-to-hard}$}
\label{sec3.1}
In Galactic XRBs, the transition luminosity from the soft to the hard
state during the decay phase spans a very narrow range of values
$\sim$ 1\%--4\% $L_ \mathrm{Edd}$ \citep{Maccarone2003,Kalemci2013}.  The average
soft-to-hard state transition luminosities of HLX-1 is $\approx 1.2
\times 10^{41}$ erg~s$^{-1}$ in the 0.3--10 keV band, with a standard
deviation of $0.6 \times 10^{41}$ erg~s$^{-1}$. We used the same
bolometric correction in \citet{Maccarone2003} to convert from 0.3--10
keV to bolometric luminosity. If the soft-to-hard state transition
luminosity of HLX-1 is at $2\%~ L_\mathrm{Edd}$ by analogy to the Galactic XRBs, the mass of the
accreting compact object is $(8\pm4) \times 10^{4} M_{\odot}$.  This
is consistent with the IMBH mass of $\sim$$10^{4}$--$10^{5}M_{\odot}$
derived from detailed X-ray spectral modelling \citep{Farrell2010, Davis2011,
  Servillat2011,  Godet2012a, Webb2012}.

\subsection{Sub-Eddington State Transitions or Super-Eddington Geometric Beaming?}

The crucial unsolved problem is whether the spectral state evolution supports 
the sub-Eddington IMBH scenario or the super-Eddington stellar-mass BH scenario.
We address this question in two steps: firstly, we test whether our results 
are consistent with an IMBH, second, we discuss whether the super-Eddington 
scenario provides a more plausible explanation.

Our spectral analysis confirmed that HLX-1 exhibits two distinct X-ray spectral states: 
the low/hard and high/soft states \citep{Godet2009,Servillat2011}. In each outburst, HLX-1 evolved 
from the hard state to the soft state during the rise phase and then 
returned to the hard state during the decay phase (\autoref{hid}). 
This spectral state evolution is similar to that of Galactic XRBs. 
HLX-1 follows the correlation between the hard-to-soft transition luminosity and the outburst peak luminosity 
which we have found  in Galactic XRBs (\autoref{h2s}). 
This correlation, which spans six orders of magnitude, strongly suggests that the same mechanism determines
the hard-to-soft state transition luminosity during an outburst for both Galactic XRBs and HLX-1. 
The X-ray spectrum in the hard state of
HLX-1 is dominated by a power-law component with a photon index of $\approx$1.6 (\autoref{fit}), 
which is similar to that of Galactic XRBs in the hard state. 
The X-ray spectrum in the soft state is dominated by a thermal component (\autoref{fit}), 
which is also similar to that of the soft state of
Galactic XRBs but with lower temperature and larger radius, as expected for an IMBH. 
The luminosity and peak color temperature 
of HLX-1 in the soft state roughly follow the $L \propto T_{\rm in}^4$ correlation, 
consistent with the prediction of the standard disk 
model and with what is observed in stellar-mass XRBs \citep{Servillat2011}. 
In addition, the larger radius and lower temperature 
(compared with stellar-mass XRBs) are self-consistent with each other and with the increase in luminosity: 
that is, both scale as expected 
for a standard disk when the BH mass is increased by a factor of $\sim$1000 to 
explain the correspondingly higher luminosity of the soft state.
We conclude that all spectral evidence is consistent with the standard disk scenario, 
that is, with sub-Eddington, not strongly beamed 
emission, in which case HLX-1 must be an IMBH. 

Disk emission is not the only way to produce a thermal X-ray spectrum. 
Optically thick outflows caused by super-Eddington accretion
may also produce a quasi-thermal X-ray spectrum with a low temperature and large radius, 
and high apparent (non-isotropic) luminosity 
if observed face-on \citep{King2003,Begelman2006,Fabrika2007,Poutanen2007,King2009}. 
The reason for this is that an optically thick outflow downscatters photons and lets them escape 
only through a collimating funnel along the polar axis (geometric beaming). 
The beaming factor $b$ in the apparent luminosity ($L = bL_\mathrm{int}$, 
where $L_\mathrm{int}$ is the intrinsic luminosity) for a face-on observer depends 
on the opening angle $\theta$ of the funnel: for a small angle, $b \approx 2/\theta^{2}$. 
Following the model of \citet{King2009}, $b \sim \dot{m}^2/73$ 
and therefore $\theta \sim 12/\dot{m}$ rad, where $\dot{m} = 0.1c^{2}\dot{M}/L_\mathrm{Edd}$. 
The intrinsic luminosity for a super-Eddington accretor 
scales as $L_{\rm int} \approx L_{\rm Edd} (1+0.6 \ln \dot{m})$ \citep{Shakura1973,Poutanen2007}. 
Combining the two factors, a face-on observer would measure an apparent luminosity 
$L \sim \left(L_{\rm Edd}/73\right)\, \dot{m}^2 \, (1+0.6 \ln \dot{m})$. Thus, if the beaming scale holds, 
a stellar-mass BH with $M \la 20 M_{\odot}$ accreting at $\dot{m} \sim 100$ would 
be seen at $L \sim 10^{42}$ erg s$^{-1}$ by an observer looking down 
the collimating funnel \citep{King2014,Lasota2015}. 
There is no physical reason preventing $\dot{m}$ from reaching values 
$\sim$100 or even much higher \citep{Wiktorowicz2015}. 
The question is whether the opening angle of the funnel continues 
to scale as $\dot{m}^{-1}$ at such high accretion rates. In the model of \citet{King2009}, 
$\dot{m} \sim 100$ corresponds 
to a funnel opening angle $\theta \sim 7^{\circ}$. This is narrower than the opening angle predicted 
by MHD simulations for comparably high accretion rates: for example, 
it is $\approx 20^{\circ}$ in \citet{Ohsuga2011} and $\approx 30^{\circ}$ in \citet{Jiang2014}. 
Those simulations suggest that at the highest accretion rates, the opening angle 
and the geometric beaming factor have a weaker dependence on $\dot{m}$ or reach an asymptotic value. 
In addition, when Compton downscattering is also properly accounted for, 
the luminosity of the photons emerging from the collimating 
funnel is less than expected simply from the beaming factor 
(that is, a larger part of the intrinsic luminosity is transferred 
to cooler electrons in the outflow). As a result, the models of \citet{Ohsuga2011} predict that the geometrically 
beamed apparent luminosity may reach only $\approx22\,L_{\rm Edd}$; 
\citet{Kawashima2012} predict peak apparent 
luminosities $\approx3 \times 10^{40}$ erg s$^{-1}$ from a 10 $M_{\odot}$ BH. 
Both values are much less than required to explain the high state of HLX-1. 
Therefore, we argue that the possibility of reaching apparent luminosities $\approx10^{42}$ 
erg s$^{-1}$ via geometric beaming of a super-Eddington stellar-mass BH is not ruled out but appears unlikely 
based on the MHD simulations \citep[\textit{e.g.}][]{Ohsuga2011,Kawashima2012,Jiang2014}, and has never been observationally proven in any real system, either.

A second difficulty of the hyper-accretion scenario is that the spectra of the photon emission predicted to 
emerge from the funnel \citep{Kawashima2012}, at least for accretion rates up to $\dot{m} = 100$, 
do not look like the 0.2 keV thermal emission seen in HLX-1. 
It is true that the emerging spectrum is predicted to soften at higher accretion rates, 
but it retains a dominant Comptonized component up to $\sim 10$ keV, 
with a roll-over at energies of $\ga 5$ keV \citep[Figures~2 and 4 in][]{Kawashima2012}, 
similar to the spectrum seen in ``normal" ULXs (making them perfectly consistent with 
the super-Eddington scenario). To our knowledge, so far, simulations have not explored what happens to the spectrum at $\dot{m} \gg 100$. 
Moreover, neither analytic slim-disk models \citep[\textit{e.g.}][]{Watarai2001} 
nor MHD simulations \citep[\textit{e.g.}][] {Kawashima2012}
predict a spectral state transition between a hard, power-law-like, 
moderately super-Eddington state, and a purely thermal, 
extremely super-Eddington regime, as would be the case in HLX-1. 
(Recall that if we interpret the high/soft state of HLX-1 
as due to geometric beaming at $\dot{m} \approx 100$, then the low/hard state would also be super-Eddington, 
with $\dot{m} \approx 30$.)
Again, there is no physical argument ruling out this possibility, but it has never been observed in real systems 
or predicted in simulations.

Another clue comes from the brightness of the optical counterpart, 
in particular, for its near-UV and blue emission. 
Early suggestions that this emission was from a massive, 
young star cluster \citep{Farrell2012} have been disproved 
after it was found that the optical emission varies substantially 
between high and low X-ray states \citep{Webb2014a}.  
For example, in the near-UV and U bands, the optical flux is $\approx 1.5$ 
mag higher in outburst than in quiescence (R. Soria et al. 2015,  in preparation). 
Thus, the most plausible interpretation is that at least the near-UV and blue emission are dominated 
by an irradiated accretion disk \citep{Soria2012}. The need for irradiation is crucial: 
a simple, non-irradiated Shakura-Sunyaev
disk is too cold and dim to fit the observed optical/UV spectrum. 
The disk must intercept and reprocess an X-ray luminosity of 
a few times $10^{39}$ erg s$^{-1}$ \citep{Soria2012, Farrell2014}. For any plausible solid angle 
subtended by the disk, this requires that the intrinsic X-ray luminosity 
in outburst be $\approx$$10^{42}$ erg s$^{-1}$, 
and that it be approximately isotropic. Very few irradiation effects on the disk would be expected 
if the observed X-ray flux emerged from a collimated funnel perpendicular to the disk plane 
and was confined by an optically thick outflow.

Taking into account all of the arguments and observational constraints outlined above, 
we conclude that at this stage, the sub-Eddington IMBH scenario with canonical state transitions 
similar to the Galactic XRBs is the {\it {least implausible}} model.

\subsection{Disk Instability Model and Comparison with Galactic LMXBTs}
\label{sec3.3}
One of the objectives of this work is to quantify to what degree HLX-1
outbursts are similar to Galactic LMXBTs outbursts. The outbursts of
bright Galactic LMXBTs are usually modeled as thermal-viscous disk
instabilities \citep[][for a review]{Lasota2001}. The correlation
between the total radiated energy and the $e$-folding rise or decay timescale,
and between the total radiated energy and the peak luminosity, found in
Galactic LMXBTs \citep{Yan2015}, are consistent with the predictions
of this model. Instead, the $e$-folding rise and decay timescales of
HLX-1 are about two orders of magnitude smaller than the extrapolation
of those correlations to a distance of 95 Mpc (\autoref{etau}). This
is consistent with the argument \citep{Lasota2011,Lasota2015} that the
thermal-viscous disk instability model cannot explain the short
outburst duration in HLX-1.

For the sake of argument, let us assume instead that we do not have any information on the
distance to HLX-1,  and let us try to constrain its distance according to the best-fitting results 
($\log E = A+B\times \log \tau$) of the $E$--$\tau$ correlations in bright 
Galactic LMXBTs \citep{Yan2015}. We would obtain a distance of 540 kpc. 
Since the intrinsic scatters of those $E$--$\tau$ correlations are quite large 
($\approx$0.4 dex), the uncertainty of the inferred distance can be up to a factor of 1.6.  
Other correlations we found in Galactic LMXBTs would be satisfied if HLX-1 were 
at a distance of 540 kpc (\autoref{ldot}, \autoref{epeak} and \autoref{etau}). 
With such a distance, we estimated the orbital period of HLX-1 ($\approx$20 hr) 
according to the correlation between optical luminosity and X-ray luminosity 
and the orbital period found in Galactic LMXBs \citep{van-Paradijs1994}, 
where the V-band magnitude of HLX-1 near the 2012 outburst peak is 
from \citet{Webb2014a}, and we estimated the mass transfer rate according to the method 
in \citet{Coriat2012}. The mass transfer rate and orbital period satisfy the criteria for a
 transient LMXB predicted by the thermal-viscous disk instability model 
\citep{VanParadijs1996,Coriat2012}.  In other words, the temporal properties of 
the HLX-1 outbursts are consistent with those of a stellar-mass X-ray transient 
at several hundreds of kiloparsecs, which can be interpreted by the thermal-viscous disk instability model. 
\citet{Lasota2015} derived a  relation between the peak luminosity and the decay timescale from 
the thermal-viscous disk instability model. According to the relation, the outbursts of HLX-1 with IMBH at 95 Mpc 
cannot be triggered by the thermal-viscous disk instability, and the thermal-viscous disk instability 
model requires a stellar-mass accretor at a distance of less than 1 Mpc.

This does {\it not} mean that HLX-1 is indistinguishable from a stellar-mass LMXBTs
hypothetically located at that distance: other outburst properties of
HLX-1 are very different. For example, the inner-disk temperature of
HLX-1 in the soft state is $\approx$0.2 keV (a distance-independent
measurement), empirically and theoretically inconsistent with the disk
temperatures of Galactic LMXBTs in the same state ($kT \sim 1$ keV). 
Moreover, the fitted inner-disk radius of HLX-1 at
95 Mpc is $\approx$50,000--100,000 km depending on inclination angle,
BH spin, and other detailed disk model assumptions. If HLX-1
were at a distance of 540 kpc, then the fitted inner-disk radius would be $\approx$250--500 km. 
Once again, this value is
inconsistent with the typical inner-disk radii of Galactic NS 
($R_{\rm in} \approx 15$ km) and  BH LMXBTs ($R_{\rm in}
\approx 50$--100 km). It is still consistent with the
inner-most stable orbit of more massive stellar BHs ($M \approx
30$--$60 M_{\odot}$). However, if the BH were in such mass range, 
and HLX-1 was at $\sim$540 kpc, then its bolometric  transition luminosity 
$L_\mathrm{soft-to-hard} would be \sim 10^{-3} L_\mathrm{Edd}$. 
This is inconsistent with the canonical state transition luminosity of BHs, 
for which $L_\mathrm{soft-to-hard} \sim 1\%$ -- $4\% L_\mathrm{Edd}$.

In summary, although the outburst timescales are surprisingly short
for an IMBH at 95 Mpc \citep{Lasota2011, Lasota2015}, and would instead indicate a stellar-mass source
at 540 kpc, this is not a viable solution for HLX-1 because at that
lower distance, its luminosity and peak temperature would no longer be
self-consistent for the disk-dominated high/soft state. In addition,
we do have other independent evidence that the distance to HLX-1 is
$\approx$100 Mpc: an optical emission line consistent with H${\alpha}$
redshifted to $\sim 7000$ km s$^{-1}$, and its projected location in
the sky in the middle of a cluster of galaxies at the same distance
\citep{Wiersema2010, Soria2013}. Conversely, we have
indirect evidence to disfavor a location at 540 kpc, because there
are no known Local Group stellar structures (dwarf or satellite
galaxies, or star clusters) in that direction of the sky at that
distance from the Milky Way. Therefore, we conclude that the
anomalously short outburst timescales indicate that either the
outbursts of HLX-1 with IMBH are not triggered by the thermal-viscous disk
instability, which is consistent with the conclusions in \citet{Lasota2011,Lasota2015}, or that 
 the efficiency of the angular momentum transport in such a disk would have to be two orders of magnitude greater than in a standard disk, or that the circularization radius and outer radius of the accretion disk are two
orders of magnitude smaller than naively estimated based on the size
of the primary Roche lobe in a circular binary system.

\subsection{Comparison with Galactic Transient Be/XRBs}

 High mass donor stars have been confirmed in some
ULXs such as NGC 7793 P13 \citep{Motch2011,Motch2014}
and M101 ULX-1 \citep{Liu2009,Liu2013}.
Therefore, it is also interesting to compare the temporal
and spectral outburst properties of HLX-1 with those of 
Galactic transient high mass X-ray binaries
(HMXBs). Be/XRBs are the most common type 
of transient HMXBs: they consist of a NS bound 
to a Be star in an eccentric orbit. This kind of systems usually
shows two types of X-ray outbursts \citep[][for reviews]{Paul2011, Reig2011}. 
Type I outbursts are quasi-periodic, and likely triggered by 
the periastron passage of the NS. The amplitude
of this type of outburst is smaller and the duration
is shorter than for the HLX-1 outbursts; besides, we have 
already noted that the periastron-passage model for HLX-1 
has serious difficulties. Type II outbursts usually show much larger 
amplitudes and longer durations, comparable with
the outbursts of HLX-1. It has been suggested 
\citep{Reig2006,Reig2008,Becker2012, Reig2013} that such outbursts include a transition 
between distinct accretion states. In an HID, transient HMXBs 
move between different states either along a horizontal branch, 
at low luminosities, or along a diagonal branch, at higher luminosities 
\citep{Reig2013}. On the diagonal branch, softer colors 
correspond to higher luminosities, as in Galactic LMXBTs and HLX-1.
The physical mechanism is still unclear, but it is not due 
to a disk instability; more likely, it depends on how the accretion 
flow is decelerated in the accretion column, and therefore 
on the balance between the accretion rate and magnetic field 
of the accreting NS \citep{Reig2013}. 
As such, any phenomenological similarities between such outbursts 
and those of HLX-1 are most likely a coincidence.

\section{CONCLUSIONS}
We studied the X-ray outburst properties of
HLX-1, using the full set of {\it Swift}/XRT monitoring observations
available to-date, which cover six outbursts from 2009 to 2015 (plus a
few sparse data points for the 2008 outburst). HIDs confirm the presence of spectral state transitions during the
outbursts \citep{Godet2009,Servillat2011}, similar to the cycle of
low/hard and high/soft states in Galactic XRBs. We found that HLX-1
follows the nearly linear correlation between
$L_\mathrm{hard-to-soft}$ and $L_\mathrm{peak}$ (\autoref{h2s}) 
and the nearly linear correlation between $\dot{L}_\mathrm{rise}$ 
and $L_\mathrm{peak}$ (upper panel of \autoref{ldot}), and that the rise timescale is similar
to those observed in stellar-mass Galactic sources; as a result, the
hard-to-soft state transition also occurs at similar times after the
beginning of an outburst. The similarity of the rise and decay
timescales and outburst duration between HLX-1 at 95 Mpc and Galactic
stellar-mass BHs remains an unexplained puzzle: on the one
hand, it suggests that the characteristic size of the accretion flow
is similar; on the other hand, we find no reason to dispute that the
accretion rate and the BH mass (and hence, the size of the
ISCO) of HLX-1 are three orders of
magnitude higher than in Galactic BHs.

For Galactic LMXBTs, the correlations between the radiated energy and the
$e$-folding rise and decay timescales, and between the radiated energy and the
peak luminosity \citep{Yan2015} quantitatively support the
thermal-viscous disk instability model. HLX-1 does not follow those
correlations, which suggests that its outbursts are not triggered by
the thermal-viscous disk instability  \citep{Lasota2011, Lasota2015}. If they were, the outer radius
of the accretion disk would have to be two orders of magnitude larger
than previously estimated, in order for a cold neutral zone to be
present \citep{Lasota2011}; but in that case, the viscosity parameter
would have to be two orders of magnitude higher than in Galactic
LMXBTs, to obtain the same (short) rise and decay timescales, 
and yet such high viscosity seems not plausible in the reality \citep{King2007}.

We quantified the evolutionary trends of the main outburst parameters
of HLX-1 between 2009 and 2015. The outburst duration, total radiated
energy, and $e$-folding decay timescale decrease along the outburst
sequence, confirming previous results
\citep{Godet2013,Miller2014}. The waiting time, recurrence time, and $e$-folding rise
timescale increase with the outburst sequence.  The increase in
the recurrence time rules out
earlier suggestions that the recurrence time corresponded to the
binary period, or indicates that the binary period increases 
in some extreme cases \citep[\textit{e.g.}][]{Godet2014, Godet2015}.
The increase of the
waiting time and the decrease of the outburst duration show that HLX-1
spends more and more time in the quiescence phase and less and less time in the outburst phase; 
however, its peak luminosity continues to reach similar values of $\approx 10^{42}$ erg s$^{-1}$. 

\acknowledgments

This work was supported in part by the National Natural Science
Foundation of China under grant No. 11333005 and
11350110498, by Strategic Priority Research Program \textquotedblleft The Emergence of
Cosmological Structures\textquotedblright under grant No. XDB09000000, and the XTP
project under grant No. XDA04060604, by the Shanghai Astronomical
Observatory Key Project. Z.Y.
acknowledges the support from the Knowledge Innovation Program of the
Chinese Academy of Sciences and National Natural Science
Foundation of China under grant No. 11403074. R.S. acknowledges an Australian
Research Council’s Discovery Projects funding scheme (project No.
DP 120102393). D.A. acknowledges support from the Royal
Society. This research made use of Astropy, a community-developed core
Python package for Astronomy\citep{Astropy-Collaboration2013}. This
work made use of data supplied by the UK {\it Swift} Science Data
Centre at the University of Leicester.  This research has made use of
the MAXI data provided by RIKEN, JAXA and the MAXI team, 
and the ASM data provided by the ASM/$RXTE$ team.


\end{document}